\documentclass[twocolumn,showpacs,preprintnumbers,amsmath,amssymb]{revtex4}

\usepackage{graphicx}
\usepackage{dcolumn}
\usepackage{bm}
\newcommand{\bi}[1]{\mbox{\boldmath$#1$}}
\newcommand{\av}[1]{\langle{#1}\rangle}
\def\be{\begin{equation}}
\def\en{\end{equation}} 
\def\p{\partial }  
\def\ve{\varepsilon}
\def\gs{\gtrsim}
\def\ls{\lesssim}

\def\bea{\begin{eqnarray}}
\def\ena{\end{eqnarray}}

\renewcommand{\theequation}{\arabic{section}.\arabic{equation}}

\begin{document}
\preprint{APS}
\title{Charged colloids 
in an aqueous mixture with a salt}

\author{Ryuichi Okamoto}
\author{Akira Onuki}
\affiliation{Department of Physics, Kyoto University, Kyoto 606-8502, Japan}

\date{\today}

\begin{abstract}
We calculate  the ion and composition 
distributions around  colloid particles 
in an aqueous mixture,  accounting for 
 preferential adsorption, 
 electrostatic interaction, 
 selective solvation  among 
ions  and  polar  molecules,  and 
 composition-dependent  ionization. 
On the colloid surface, 
we predict a precipitation  transition 
induced by strong preference 
of hydrophilic ions to water 
and a prewetting  transition 
between weak and strong  adsorption and ionization. 
These transition lines 
extend far from the solvent coexistence curve 
in the plane of  the interaction parameter $\chi$ 
(or  the temperature)  and the 
average solvent composition. 
The colloid interaction 
is drastically altered  by 
these phase transitions on the surface. 
In particular, the  interaction 
 is much amplified upon bridging of 
wetting layers   formed 
above the precipitation line.  
Such wetting layers 
can either completely or partially cover 
the colloid surface  depending   on the 
average solvent composition. 
\end{abstract}

\maketitle

%
%

\section{introduction}

Extensive efforts have been made to understand 
the interaction among ionized colloid particles 
in a solvent \cite{Dej,Ov,Russel,Ohshima,LevinReview}, because they form 
model crystal and glass at high densities. 
Recently, considerable  attention has also been paid 
to the effect of  preferential  adsorption 
of one of the components in a mixture solvent.  
Several groups \cite{Beysens,Maher,Kaler,Guo,Bonn}   observed  
aggregation of colloidal particles near the 
coexistence curve  in one-phase states of  a 
binary mixture of 2,6-lutidine and water.
 For small ionization 
 the adsorption of lutidine  
was increased in water-rich states, 
while for large  ionization 
that of  water   was increased in lutidine-rich states. 
 This means  that the colloid surface 
can either repel or attract water,  
depending on the degree of ionization.  
It is worth noting that   polyelectrolytes 
are often hydrophobic without  ionization, 
but becomes effectively hydrophilic even at 
low ionization \cite{Barrat,Holm,Rubinstein}.  

At high colloid concentrations, 
a  flocculated phase 
rich in colloids emerges  
\cite{Beysens}, which  
changes from gas, liquid, fcc crystal, and glass 
with increasing the colloid concentration \cite{Guo}. 
Such aggregation was  claimed 
to be a result of a  true phase separation in ternary 
mixtures \cite{Kaler}. 
A microscopic theory by  Hopkins {\it et al.} 
\cite{Evans} indicated  the aggregation mechanism. 
They treated  neutral   colloids 
 coated  by a   thick adsorption  layer 
rich in the preferred 
component  in one-phase environments 
 rich in  the other component. 
 They found that this adsorption is intensified 
near the coexistence curve, strongly influencing  
the colloid interaction.

In  understanding  these experiments, 
however,  attention has not been paid 
to the selective solvation (hydration 
for aqueous mixtures) among 
charged particles (ions and ionized parts on the 
colloid surface) and polar solvent 
molecules \cite{Is}. The solvation  effects    
 have not yet been adequately 
 investigated in soft materials, but 
 they can  influence  the phase 
 separation  behavior  profoundly 
 and even give rise to a  
new phase transition \cite{Onuki-Kitamura,OnukiPRE,Nara,Current}.  
 The solvation  
chemical potential  $\mu_{\rm sol}(\phi)$ 
of a hydrophilic ion stems from the ion-dipole 
interaction and strongly depends 
on the ambient composition of water.  
In  electrochemistry, 
 the following chemical potential 
 difference has been measured  \cite{Hung,Osakai}:  
\be 
\Delta\mu_{\rm sol}^{\alpha\beta}= 
\mu_{\rm sol}(\phi_\beta)-\mu_{\rm sol}(\phi_\alpha), 
\en  
between two coexisting phases,  where  
$\phi_\alpha$ and $\phi_\beta$ 
are the bulk water compositions in the two phases. 
This quantity  is the so-called Gibbs transfer 
free energy per ion, which  
determines the ion partition 
and a Galvani potential difference   between the two phases. 
Its  magnitude typically 
much exceeds  $k_BT$ (about  
$15k_BT$ for Na$^+$ and  Cl$^-$ 
in water-nitrobenzene at room temperatures).  
 Furthermore, the dissociation  of ionizable groups 
on the colloid surface should be treated  as a  chemical reaction 
sensitively  depending  on the local environment 
(the solvent composition and the local 
electric potential)  as in the case of polyelectrolytes 
\cite{Barrat,Holm,Bu2,Onuki-Okamoto}. 
Therefore, even  very small composition variations 
 around the colloid surface 
can induce  significant changes in the ion distribution, 
 the electric potential, and the degree of ionization, 
so it can  drastically alter   the colloid interaction.

Historically,  the colloid interaction 
has been  supposed to consist of the 
screened Coulomb repulsive 
infraction  $F_{\rm DLVO}$ and the van der Waals 
attractive interaction  $F_{\rm vdw}$ 
since    the celebrated theory 
developed   by   Derjaguin, Landau, 
Verway, and Overbeek (DLVO) \cite{Russel,LevinReview,Dej,Ov}.  
For two colloid particles  with radius $a$, 
the former reads 
\be 
F_{\rm DLVO}=  
\frac{{\bar Q}^2e^{-\kappa(d-2a)}}{\bar{\ve}(1+ \kappa a)^2d}
\en 
where $d$ is the distance between two colloid centers, 
$\bar Q$ is the average charge 
on a colloid particle, $\bar \epsilon$ is the average 
dielectric constant (at the average composition 
for a mixture  solvent), and  
$\kappa$ is the Debye wave number.  
The latter  arises from 
the pairwise van der Waals  interaction 
($\propto -1/r^6$) among constituent molecules, where 
 $r$ is the distance between two molecules. 
In terms of the Hamaker constant $A_{\rm H}$, $F_{\rm vdw}$  between 
two identical 
colloids with radius $a$ is written as \cite{Russel}  
\be 
F_{\rm vdw}
= -\frac{A_H}{6} \bigg [ \frac{2a^2}{d^2-4a^2} +\frac{2a^2}{d^2} 
+ \ln \bigg(1- \frac{4a^2}{d^2}\bigg)\bigg] .
\en 
It  grows  for  short separation  $\ell=d-2a \ll a$  as 
\be 
F_{\rm vdw}  
\cong -{A_{\rm H}}a/{12}{\ell}  ,
\en 
while  it decays as 
$ - {16 A_{\rm H }}a^6/{9}{d^6}$
for long separation  $\ell \gg a$.  
 We assume that $F_{\rm vdw}(d)$ should 
  saturate  at a short distance  on the 
order of the solvent molecular diameter $a_0$ 
($\sim 3{\rm \AA}$).  Then  $F_{\rm vdw}(d)$ 
decreases  down to  $-A_H a/12a_0$ at  contact.
The size of $A_H$ strongly depends on binary mixtures 
and colloid particles under investigation 
and can be made small by matching the dielectric 
constants of the colloid and the solvent. 
In this paper, $F_{\rm vdw}$ will  not be included 
in our theoretical scheme for simplicity.

In  binary mixtures, 
the adsorption-induced 
composition disturbances 
  give rise to an attractive interaction  
$F_{\rm ad}$   between two colloid particles  \cite{Beysens}. 
A linear theory can then be developed 
when the adsorption and the ionization are both  weak.  
Further in the special case of 
weak selective solvation, the pair interaction is 
the sum of $F_{\rm DLVO}$ in Eq.(1.2)  
 and $F_{\rm ad}$  given by  
\be 
F_{\rm ad}= 
- A_{\rm ad}a^4 \xi^2  \frac{ 
e^{-\ell /{\xi}}}{(a+ {\xi})^2d} ,     
\en 
where $\xi$ is  the correlation length 
 growing near the solvent criticality. 
 The coefficient  $A_{\rm ad}$ is 
proportional to $ h_1^2$, where $h_1$ is 
the surface field arising from 
 preferential  molecular interactions 
between the surface and the two solvent species 
\cite{Cahn,Binderreview,Bonnreview}. 
In deriving Eq.(1.5),  
$h_1$ is assumed to be small 
and Eq.(1.5) is invalid very close to the criticality \cite{h1}; 
nevertheless, 
$F_{\rm ad}$ becomes  the dominant interaction 
at long distances for $\xi> \kappa^{-1}$.
We shall see that
  $F_{\rm ad}$  can exceed  
 $F_{\rm vdw}$ even at the molecular separation  
$ a_0$  for small $A_H$.    Furthermore, on approaching 
the solvent criticality without ions, 
 the  adsorption-induced interaction  
among solid objects (plates, rods, and spheres) 
 becomes  universal (independent 
of the material parameters)   
in the limit of strong adsorption \cite{Fisher}, so it 
has been  called the critical  Casimir interaction 
 \cite{Krech,JSP,Nature2008,Gambassi}. 
However, it should be 
affected by  ions for strong 
selective solvation.


In this paper, we  aim  to investigate  the 
ion effects on the colloid interaction  
in binary mixtures using   a coarse-grained 
approach.  A merit  of our  approach is 
that we can  treat   
 the preferential solvation in its 
strong coupling limit.   
In our  recent work  \cite{Current,Okamoto}, 
we found that 
a strongly selective solute    can   induce  formation  
 of   domains rich in the preferred  component     
even  far  from  
the solvent coexistence curve.    
We shall see that this precipitation phenomenon 
can  occur on the colloid surface, 
leading to a wetting layer 
coating  the colloid surface.
As another prediction,  
there can be a first-order prewetting 
surface transition between  
weak and strong  adsorption  far from 
the solvent criticality, 
as discussed in our paper on charged rods  
\cite{Onuki-Okamoto}. 
These two  phase transitions 
 occur  when the 
volume fraction of the 
selected  component is relatively small ($\phi<\phi_c$). 
With intensified mutual interactions, 
colloid particles should trigger 
a macroscopic phase separation  
to form a floccuated phase \cite{Beysens,Guo,Kaler}.
As a closely  related  example, 
precipitation of DNA has been  observed 
with addition of  ethanol in water 
\cite{B1,B2,B3}, where  the ethanol  added  is 
excluded from condensed DNA.

In addition to usual  hydrophilic ions, 
we are also interested in the colloid interaction 
in the presence of antagonistic ion pairs 
in aqueous mixtures, 
where the cations  are hydrophilic and 
the anions are hydrophobic,  or  vice versa \cite{Nara,Current}.   
A well-known example 
in electrochemistry   is   
a pair of  Na$^-$ and 
tetraphenylborate BPh$_{4}^-$, where the latter anion 
 consists  of   four phenyl rings bonded to an ionized 
boron and acquires strong hydrophobicity.   
Such ion pairs behave 
antagonistically in the presence of composition 
heterogeneities, giving rise to  
 formation of mesophases,  
as recently observed by Sadakane {\it et al.}
by adding a small amount of NaBPh$_{4}$
to D$_{2}$O and tri-methylpyridine 
 \cite{Sadakane}.  They should produce 
an  oscillatory interaction between 
walls or colloid particles as in the case of 
liquid crystals \cite{Uchida}.

The organization of this paper is as follows. 
In Sec.II,  we will present   a  Ginzburg-Landau 
model of  a binary mixture containing ions and ionizable 
colloid particles, where the bulk part 
  includes  
the electrostatic and solvation interactions 
 and the surface part 
  the dissociation free energy. 
In Sec.III, we will examine  the linearized 
version of our theory for 
 the electrostatic and composition fluctuations 
 as  a generalization of  the Debye-H$\ddot{\rm u}$ckel 
and DLVO theories. 
In Sec.IV, we will discuss how 
 a wetting layer is formed 
on the colloid surface, which  takes  place as 
a precipitation phase transition. In Sec.V, 
we will present numerical results 
on the basis of our  nonlinear scheme, 
where we shall encounter  precipitation and 
prewetting phase transitions on the  surface 
even far from the solvent coexistence curve.
We shall also see bridging of wetting layers 
and  a changeover between complete and 
partial wetting.

\section{Ginzburg-Landau model 
for a mixture solvent}
\setcounter{equation}{0}

We suppose monovalent 
hydrophilic cations and anions 
in a binary  solvent 
composed of a water-like polar 
component (called water) and a less polar component 
(called oil) in a  cell with a  volume $V$ 
\cite{Onuki-Kitamura,OnukiPRE,Okamoto}. 
 We  also place  one or two 
 negatively ionizable 
colloid particles in the cell. 
Experimentally, we suppose a dilute suspension 
of colloid particles in a mixture solvent. 
To apply our results to such systems, we 
should set  $V=1/n_{\rm col}$, 
where $n_{\rm col}$ is  the  colloid density.  

For simplicity,  we neglect 
the van der Waals interaction 
 $F_{\rm vdw}$.  The Boltzmann constant 
$k_B$ will be set equal to unity 
in the following.

\subsection{Free energy including electrostatics, 
 solvation, and surface interaction}
We assume that 
 the counterions coming from the colloid surface 
are of the same species as the cations added 
as a salt. The cation and anion number 
densities are written as $n_1$ and $n_2$, 
respectively, while the  water composition  
is written as $\phi$. We treat these variables 
as smooth functions in space.  
The total free energy 
$F_{\rm tot}=F+F_s$ consists of the bulk part 
$F$ and the surface part $F_s$. 
The former  is written as    
\be
F = \int' d{\bi r}[f_{\rm tot} + \frac{TC}{2}|\nabla\phi|^2] 
+ \int d{\bi r} 
\frac{\varepsilon  {\bi E}^2}{8\pi } ,
\en 
where   $\int' d{\bi r}$ 
 is  the space integral  in 
  the colloid exterior in the cell 
 and  $\int d{\bi r}$ is that in the whole cell 
including the colloid interior. 
The electric potential $\Phi$ is defined 
even in the colloid interior. 
We assume that inhomogeneity of $\phi$ gives rise to 
the gradient free energy, where 
the coefficient $C$  is a positive constant.

In the first term of Eq.(2.1), the free energy density 
 $f_{\rm tot}$ is the chemical part depending   on 
$\phi$, $n_1$, and $n_2$ as   
\be
{f_{\rm tot}} = f(\phi) 
 +  T\sum_i n_i  \bigg[\ln (n_i\lambda_i^3) -1-  g_i \phi\bigg].
\en 
In this paper, the  molecular volumes of 
the two solvent components  take 
a common value $v_0$, though 
they are often  very different 
in real binary mixtures. As a  molecular length, 
we introduce  
\be 
a_0=v_0^{1/3},
\en 
which  is 
supposed to be of order $3{\rm \AA}$. Then $C 
\sim a_0^{-1}$  \cite{comment-gra}.  
The colloid radius $a$ is much larger than  $a_0$.  
(In our previous papers \cite{Onuki-Kitamura,OnukiPRE,Nara,Current,Okamoto}, 
$a$ has been used to denote the molecular length  
$v_0^{1/3}$.)  
We neglect the volume fractions of the ions 
assuming their small sizes. 
In our numerical analysis, we adopt 
the    Bragg-Williams form \cite{Onukibook},        
\be 
\frac{v_0}{T}f   =   
 \phi \ln\phi + (1-\phi)\ln (1-\phi) 
+ \chi \phi (1-\phi),   
\en 
where   $\chi=\chi(T)$ is the interaction 
parameter depending  on   $T$.  
The critical value of $\chi$ is 2 without ions. 
In the ionic part of  Eq.(2.2), 
$\lambda_i = \hbar(2\pi/m_i T)^{1/2}$ 
is   the thermal de Broglie wavelength of the species $i$ 
with $m_i$ being its  mass. 
The dimensionless parameters  
$g_1$ and $g_2$  represent  the 
degree of selective solvation 
\cite{Onuki-Kitamura,OnukiPRE}, in terms of which 
the solvation chemical potential 
is $\mu_{\rm sol}^i= {\rm const.}- Tg_i\phi$ 
and the Gibbs transfer free energy is $Tg_i (\phi_\alpha- 
\phi_\beta)$ for the ion species $i$ (see Eq.(1.1)). 
 If $\phi$ is the water composition, 
we have $g_i >0$ for  hydrophilic ions 
and $g_i <0 $ for hydrophobic ions.  
In many aqueous mixtures,  the amplitude 
$|g_i|$ well exceeds  
$10$ both for hydrophilic ions and 
hydrophobic solutes \cite{Current}.

The last term in Eq.(2.1) 
is the electrostatic part\cite{Tojo}, where 
  ${\bi E}=-\nabla\Phi$  is the electric field. 
  The   electric  potential $\Phi$  is defined 
  in the whole region  including the colloid interior. 
We assume continuity of $U$ 
through the colloid surface 
neglecting surface  molecular polarization. 
It follows    the   Poisson equation,    
\be 
\nabla\cdot{\bi D}= 
-\nabla\cdot\ve\nabla \Phi=  4\pi \rho,   
\en 
where ${\bi D}= \ve{\bi E}$ is the electric flux density. 
The  dielectric constant $\ve(\phi)$  is   
 assumed to be 
 a linear function of $\phi$ 
 in the  solvent. Thus it behaves as      
\bea 
\ve(\phi)&=&\ve_0 + 
\ve_1 \phi ~\quad ({\rm colloid~ exterior})\nonumber\\ 
 &=&\ve_{\rm p} \qquad\qquad ({\rm colloid~ interior}), 
\ena   
which  is   
  $\ve_0$ in   oil (at $\phi=0$),   
$\ve_0+\ve_1$ in    water  (at $\phi=1$), 
and $\ve_{\rm p}$ in the colloid interior. 
Let  all the charges be  monovalent. Then  
the charge density $\rho$ is written as 
\be 
\rho=e(n_1-n_2) -e \sigma \delta_s. 
\en  
The  first bulk term  is  
nonvanishing in the colloid  exterior and 
the second  part arises  from 
 the areal  density $\sigma$ of 
 the ionized groups on the colloid surface, where 
  $\delta_s$ is  the delta  function 
nonvanishing only on  the colloid surface.  
There is no charge density 
in the colloid interior. As a result, there 
arises a discontinuity in the normal component  
${\bi \nu}\cdot{\bi D}$, where 
${\bi \nu}$ is  the outward 
normal unit vector on the colloid surface. Let 
 ${\bi D}_+$ and  ${\bi D}_-$ be the 
values of ${\bi D}$ immediately outside and 
inside the colloid surface, respectively. Then, 
\be 
{\bi \nu}\cdot({\bi D}_+ - {\bi D}_- )
 =-4\pi e \sigma .
\en 
On the other hand, on the cell boundary,  
we  assume  no surface free energy and no surface charge. 
In our simulation, we   thus set   
\bea 
&&
\bi{\nu}_b \cdot\nabla\phi=0, 
\nonumber\\ 
&&
\bi{\nu}_b \cdot{\bi E}= -
\bi{\nu}_b \cdot{\nabla\Phi}=0, 
\ena 
where $\bi{\nu}_b$ is its  normal 
vector of the cell boundary.

 The density of the ionizable  groups 
 on the colloid surface is 
 written as  $\sigma _0$. 
 The fraction of ionized groups or the degree of ionization 
  $\alpha$ is defined in the range $0\le \alpha\le 1$. 
  The density of the ionized groups is written as \cite{Onuki-Okamoto}  
 \be
\sigma =\sigma _0 \alpha.
\en
We treat $\alpha$ as a 
fluctuating  variable 
depending on the local composition and potential. 
The surface  free energy 
$F_s$ depends on $\phi$ and $\alpha$ as 
\be 
F_s= \int dS (T \gamma\phi + f_d),   
\en  
where $\int dS$ is  the integration  
on the colloid  surface. Here we neglect 
the second order contribution ($\propto \phi^2$) 
present in the original theory \cite{Cahn} 
to the surface free energy density    
(though it  is relevant near 
the critical point for neutral fluids 
\cite{Binderreview}). 
The coefficient  $\gamma$ represents 
the short-range  interaction between 
the mixture solvent and the colloid surface (per 
solvent molecule)\cite{Cahn}. 
We call $\gamma$  the surface 
interaction parameter or  the surface field  
(though  $h_1 \equiv 
-T\gamma$ is usually called the  surface field 
in the literature \cite{Binderreview,Bonnreview}). 
 The $f_d$ in Eq.(2.11) is the dissociation 
 (ionization) free energy density of the form 
 \cite{Barrat,Bu2,Onuki-Okamoto},  
\be
\frac{f_d}{T\sigma_0  }=
\alpha\ln\alpha +(1-\alpha)\ln(1-\alpha)+\alpha(\Delta_0-\Delta_1\phi),
\en
where the first two terms 
arise from  the entropy of selecting the ionized  
groups among the ionizable ones, while  $\Delta _0-\Delta _1\phi$ 
is the composition-dependent  ionization free energy 
divided by $T$.  We suppose that  
 the ionization is much enhanced 
 with increasing the water content, which means that $\Delta_1$ 
 should be  considerably larger than unity.

\subsection{Equilibrium  relations}

In our finite system, the cation number 
increases with an increase of ionization, 
while the numbers of the  anions and the 
solvent  particles are fixed.  
Let  $n_0$ be 
the average   density of the added salt  
and $\bar \phi$ be  the average water composition. 
They are important parameters in our problem 
as well as $\chi$. Then, 
\bea 
&&\int' d{\bi r}(n_1({\bi r})-n_0)= \int dS\sigma, \\ 
&&\int' d{\bi r}(n_2({\bi r})-n_0)=0, \\
&& \int' d{\bi r}(\phi({\bi r})-\bar{\phi})=0.
\ena
The right hand side of Eq.(2.13) 
is  the number of the counterions 
from the colloid surface. 
These relations are 
consistent with the expression for 
the charge density $\rho$ in Eq.(2.7).  
In  equilibrium we should minimize  
the grand potential $\Omega$  defined by  
\be 
\Omega =  F- \int' \hspace{-1mm} d{\bi r} 
(h \phi  +\sum_i \mu_i n_i)+ F_s+ \int \hspace{-1mm} 
dS \mu_1 \sigma. 
\en   
Here  $h$, $\mu_1$, and $\mu_2$ are introduced 
as  Lagrange multipliers owing to the constraints (2.13)-(2.15).  
They    have   the meaning  
 of the chemical potentials expressed as  
 $h=\delta F/\delta \phi$ and 
  $\mu _i =\delta F/\delta n_i$.

To minimize $\Omega$, 
we  superimpose infinitesimal deviations 
$\delta\phi$, $\delta n_1$, $\delta n_2$, 
and $\delta \alpha$ on $\phi$, $ n_1$, $ n_2$, 
and $\alpha$, respectively. 
First, we calculate  the infinitesimal 
 variation of the electrostatic part 
 $F_e\equiv \int d{\bi r}\ve{\bi E}^2/8\pi$ in $F$ in Eq.(2.1).  
From  the 
relation   $\delta (\ve {\bi E}^2) 
= 2{\bi E}\cdot\delta{\bi D}- {\bi E}^2\delta\ve$, 
we obtain 
\bea 
\delta F_e&=&\int' d{\bi r}\bigg[\Phi\delta\rho- 
\frac{{\bi E}^2}{8\pi}\ve_1{\delta\phi}\bigg] 
-\int dS~ e\sigma_0 \delta\alpha \nonumber\\
&& 
-\int_{\rm cell} dS~ \Phi {\bi \nu}_b\cdot 
(\ve \delta{\bi E}+ \delta\ve{\bi E}) ,
\ena   
where the  integration is in the colloid exterior 
in the first term,  
on the colloid surface in the second term,  
and  on the cell boundary in  the third  term. 
 From Eq.(2.9) we have ${\bi \nu}_b\cdot {\bi E}=
{\bi \nu}_b\cdot \delta{\bi E}=0$ on the collied 
surface, so 
the third term    vanishes.
From Eqs.(2.1) and (2.17) we obtain   
\bea 
\hspace{-1cm}h=&& f'(\phi)-
TC\nabla ^2 \phi -T\sum _i g_in_i-\frac{\ve_1}{8\pi}{\bi E}^2, \\
&&\hspace{-1cm}\mu_i = T\ln(n_i\lambda_i^3) -Tg_i \phi +Z_i e\Phi, 
\ena
where  $f'=\p f/\p \phi$ in Eq.(2.18) and 
$Z_1=1$ and $Z_2=-1$ in  Eq.(2.19). 
It folows  the modified 
Poisson-Boltzmann relations  for the ion densities,  
\be
n_i= n_i^0 \exp(g_i\phi-Z_i U), 
\en 
where $n_i^0= \lambda_i^{-3} \exp({\mu_i/T})$ are constants. We 
introduce the normalized electrostatic 
 potential, 
\be 
U=e\Phi/T. 
\en

Using the above $h$ and $\mu_i$,  
we calculate the incremental changes  of  $F$  and $F_s$   as 
\bea 
&&\hspace{-2cm}
\delta F= \int' d{\bi r}[ h\delta\phi+ \sum_i \mu_i \delta n_i]
\nonumber\\
&&\hspace{-1.5cm}
-\int dS [(C\bi{\nu}\cdot\nabla\phi) \delta\phi 
+(e\Phi \sigma_0)\delta \alpha],\\
&&\hspace{-2cm}
\delta F_s=T \int dS( \gamma - \sigma\Delta_1)\delta\phi \nonumber\\
&&\hspace{-1.5cm}+T\sigma_0 \int dS \bigg[\ln \frac{\alpha}{1-\alpha}+\Delta_0-\Delta_1\phi \bigg] \delta \alpha.
\ena 
Vanishing of the surface terms 
proportional to $\delta\phi$ 
in $\delta F+ \delta F_s$ 
yields the boundary condition of $\phi$ 
on the colloid surface written as \cite{Okamoto}  
\be 
C\bi{\nu}\cdot\nabla\phi=   
\gamma - \Delta_1\sigma_0\alpha,
\en 
where $\bi\nu$ is the outward 
normal unit vector on the colloid surface. 
In the same manner, vanishing of the terms 
proportional to $\delta \alpha$ yields 
 \cite{Onuki-Okamoto}
\be 
\frac{\alpha}{1-\alpha}= 
\exp(-\Delta_0 +\Delta_1\phi+U -\mu_1/T). 
\en

 We multiply Eq.(2.25)  by 
$n_1$ in Eq.(2.20) at  the surface to obtain  
the mass action law on the surface, 
\be 
\frac{\alpha n_1}{1-\alpha}=K(\phi), 
\en 
where the factor $\exp(U-\mu_1/T)$ is cancelled.   
We introduce  the composition-dependent 
ionization  constant by   
\be 
K(\phi)=\lambda_1^{-3} 
 \exp[-\Delta_0 +(\Delta_1+g_1)\phi],
\en     
in terms of which we have $\alpha= K(\phi)/[n_1+K(\phi)]$. 
Thus we have weak ionization $\alpha\ll 1$ for 
$n_1\gg K(\phi)$ and strong 
ionization $\alpha\cong 1$ for $n_1\ll K(\phi)$ on the surface.  

In addition, from Eq.(2.25),  
the ionization free energy density $f_d$ in Eq.(2.12) 
  becomes 
\be 
f_d/T \sigma_0 = \alpha (U-\mu_1/T) +\ln (1-\alpha),  
\en 
which will be used in deriving Eq.(3.17). 

\subsection{Changeover from hydrophobic to 
hydrophilic surface with progress of ionization}

We further discuss the consequence 
of the boundary condition (2.24). 
Near the surface, oil is enriched 
for $\phi'= \bi{\nu}\cdot\nabla\phi>0$, while  
water  is enriched 
for $\phi'<0$. 
For very small $\alpha$, 
the colloid surface is   hydrophobic 
for $\gamma>0$ and is hydrophilic 
for $\gamma<0$. 
However, with increasing $\alpha$,   
 an originally hydrophobic 
surface  can become  effectively 
hydrophilic  if  
\be 
0<\gamma< \Delta_1\sigma_0,
\en  
under which  the surface derivative 
$\phi'$ becomes negative 
for $\alpha>\gamma/\Delta_1\sigma_0$. 
In contrast, if $\gamma>\Delta_1\sigma_0$, 
the surface remains hydrophobic even for $\alpha=1$. 

This  weakly hydrophobic 
 situation can well happen in real 
colloid systems in  mixture solvents 
for not small   $\sigma_0$ owing to 
strong composition-dependent ionization.
As stated in Sec.I,  
colloid aggregation in near-critical 
lutidine-water occurred at  
lutidine-rich compositions  
for small ionization and at  water-rich compositions  
for larger  ionization \cite{Beysens,Maher}. 
In our theory this means that  the colloid surface 
remained  hydrophobic for small ionization, 
while it became hydrophilic for large  ionization.  
It is also well-known that 
hydrophobic polyelectrolytes (without ionization) 
can become hydrophilic with  progress of  ionization 
\cite{Rubinstein}.

\section{Linear theory in one-phase states}
\setcounter{equation}{0}

The  Debye-H$\ddot{\rm u}$ckel 
and DLVO theories \cite{Russel,LevinReview,Dej,Ov} are justified 
for small electrostatic perturbations, where  
 the amplitude of the normalized potential 
 $U=e\Phi/T$ should be  smaller than unity. 
 Here we present  a generalized linear theory, 
 including  the composition fluctuations in 
  a mixture solvent.  To justify the linear treatment,  we  
  assume  that the degree of ionization $\alpha$ 
and the surface field  $\gamma$ are 
both very small. Treating  
$\alpha$ and $\gamma$ 
 as small  parameters,  
we   calculate $U$ and the  composition deviation, 
  \be 
  \psi=\phi-\bar{\phi}.
\en 
to  linear order in $\alpha$ or $\gamma$.   
From Eq.(2.24) the preferred component 
is only weakly adsorbed 
 on the colloid surface. 
These deviations produce a change in 
the grand potential 
$\Omega$ in Eq.(2.16)  of second order ($\propto 
\alpha^2, \alpha\gamma$, or $\gamma^2)$.

\subsection{Linearized relations}

In the limit of large cell 
volume $V\gg a^3$ or in the dilute 
limit of  colloid suspension, we may assume that 
$\phi$,  $n_i$, and $U$ tend to 
$\bar\phi$, $n_0$, and 0, respectively, 
exponentially far from the colloids. 
From Eqs.(2.18) and (2.19) 
we then have   $h=f'({\bar\phi})-T\sum_i g{\bar \phi}n_0$ 
and $\mu_i= T\ln (n_0 \lambda_i^3)-Tg_i{\bar\phi}$. 
in terms of   $\bar\phi$ and $n_0$.  
From Eq.(2.20)  we obtain 
\bea 
\delta n_1 &=&n_1-n_0 \cong 
  n_0 (g_1 \psi -U),\nonumber\\
\delta n_2&=&n_2-n_0 \cong n_0 ( g_2 \psi +U).
\ena 
From  Eq.(2.25) the expansion of $\alpha$  is of the form,  
\be 
\alpha \cong  \bar{\alpha} [1+\Delta_1 \psi+U],
\en  
where the surface values of $\psi$ and $U$ are used 
and $\bar{\alpha}= K(\bar{\phi})/n_0$ 
is the degree of ionization in the 
homogeneous case  
assumed  to be very small.   
The deviation $\delta \alpha=\alpha-{\bar \alpha}$ 
is already of second order. 
We shall see that $\delta \alpha$  gives  rise to 
a third order contribution to $\Omega$ 
 and may be neglected in our linear theory. 
 The colloid surface is under  
 the fixed charge condition in the linear theory. 
We should calculate  $U$ in the whole space 
imposing   $\nabla^2 U=0$ in the colloid interior, while 
 $\psi$, $n_1$, and $n_2$ are  
 defined only in  the  colloid exterior.

In terms of 
the average dielectric constant ${\bar\ve}= \ve(\bar{\phi})$, 
we  introduce  the Bjerrum length $\ell_B$ and the 
Debye wave number  $\kappa$    by 
\bea 
&&\ell_B= e^2/\bar{\ve}T, \\
&&\kappa= (8\pi n_0e^2/\bar{\ve}T)^{1/2}=
(8\pi \ell_Bn_0)^{1/2}.  
\ena  
 Without coupling to the ion densities, 
 the correlation length 
of the composition fluctuations is given by 
 $\xi = (C/\tau)^{1/2}$ in one-phase 
states  with  
\be 
 \tau=\frac{1}{T}f''({\bar\phi}) 
=\frac{1}{v_0}[ \frac{1}{{\bar\phi}(1-{\bar\phi})}-2\chi], 
\en 
where $f''(\phi)= \p^2 f/\p \phi^2$ and use is  made 
of Eq.(2.4).

In the bulk region of the colloid exterior, 
Eqs.(2.5) and (2.18) are linearized 
with respect to  $\psi$ and $U$ as  
\bea 
&&\hspace{-0.4cm} \nabla^2 U= 
\kappa^2( U- g_a \psi),\\
&&\hspace{-1.4cm}C\nabla^2\psi= 
( \tau - \tau_c )\psi
 + 2n_0g_a(U-g_a\psi),
\ena 
where   we introduce two coefficients,
\bea
&&\hspace{-1cm} \tau_c=
 n_0 (g_1+g_2)^2/2,  \\
&&\hspace{-1cm}g_a= (g_1-g_2)/2.
\ena  
As can be seen from the structure factor 
of the composition in  Appendix A, 
  $\tau_c$ is the  shift of the spinodal 
in the long wavelength limit 
due to the selective solvation. 
The size of $\tau_c$  can be significant even for small $n_0$ 
for  $g_1 \gg 1$ and $g_2\gg 1$,  
 which agrees  with experimental  large 
shits of the coexistence 
curve induced by  hydrophilic ions \cite{polar1}. 
We assume  $\tau>\tau_c$ to ensure the thermodynamic 
stability. The correlation length 
 of  $\psi$ is changed by ions as  
\be 
{\bar\xi}= [(\tau-\tau_c)/C]^{-1/2},  
\en 
which grows as $\tau\to \tau_c$ 
or as $\chi \to \chi_s\equiv  
1/2{\bar\phi}(1-{\bar\phi})-v_0\tau_c/2$. 
The  $g_a$  arises from the asymmetry of the selective 
solvation  between the cations and the anions,  
giving rise to the coupling of  $U$ and $\psi$.

On the colloid surface,  Eqs.(2.8) and (2.24) yield   
the linearized boundary conditions, 
\bea 
&&{\bar\ve}({\bi \nu}\cdot \nabla U)_+- 
{\ve_p}({\bi \nu}\cdot \nabla U)_-= 4\pi e^2 \bar{\sigma}/T,\\
&&({\bi \nu}\cdot \nabla \psi)_+ = \bar{\gamma}/C, 
\ena
where  $(\cdots)_+$ and $(\cdots)_-$ denote 
taking the values immediately outside and inside the 
colloid surface, respectively. The 
$\bar\sigma$ and $\bar\gamma$ 
are averages defined by 
\bea 
&& \bar{\sigma}=\sigma_0\bar{\alpha}, \\
&&\bar{\gamma}= \gamma -\Delta_1 \bar{\sigma}. 
\ena 
In $f_d$ in Eq.(2.28) we use the expansion  
$\ln(1-\alpha)= -\alpha+ 
\alpha^2/2+\cdots $, where $\alpha$ 
behaves as in  Eq.(3.3) and $\alpha^2 \cong 
{\bar\alpha}^2$. 
Up to the second order we  find  
\be 
f_d/T\sigma_0 =-\alpha \mu_1/T -{\bar\alpha}
- {\bar{\alpha}} \Delta_1 \psi 
   + {\bar\alpha}^2/2.  
\en 
Thus the last two terms  in the grand potential 
 $\Omega$ in Eq.(2.16)  become $F_s+\int
 dS \mu_1\sigma = T \int dS 
 {\bar\gamma}\psi +$const. in terms of $\bar\gamma$, 
 where the first term in the right hand side of Eq.(3.16) 
cancels to vanish.  Hence the  second-order contributions to 
$\Omega$  are  written as  
\bea 
\frac{\Delta \Omega}{T}
  \hspace{-1mm}& =& 
   \hspace{-1mm}\int' \hspace{-1mm} d{\bi r}\bigg[\frac{\tau}{2}\psi^2 + 
 \frac{C}{2}|\nabla\psi|^2+
 \sum_i \bigg(\frac{\delta n_i^2}{2n_0}-
 g_i\psi \delta n_i\bigg) \bigg]\nonumber\\
&&  
+ \int \hspace{-1mm}d{\bi r}
{\bar{\ve} {\bi E}^2}/{8\pi T}+
 \int \hspace{-1mm}dS \bar{\gamma}\psi\nonumber\\ 
&=&  \frac{1}{2}\int dS(\bar{\gamma} \psi- {\bar \sigma}U),
\ena     
 where  the bulk integrations in the first two lines  
are transformed   into the surface ones in the third line 
with the aid of Eqs.(3.7) and (3.8). 
With the third line, we thus  need to calculate 
only the surface averages of $\psi$ and $U$ at fixed 
surface charge in the linear theory.

\subsection{Two characteristic wave numbers $q_1$ and $q_2$}

In the colloid exterior, 
there arise   two characteristic 
wave numbers, denoted by $q_1$ and $q_2$.  
  If  $g_a = 0$, 
$U$ varies  on the  scale of 
 the Debye screening length  $\kappa^{-1}$, 
 where $\kappa$ is defined  in Eq.(3.5),  and 
$\psi$ varies on the scale of 
$\bar\xi$ in Eq.(3.11).  For $g_a \neq 0$, 
they  are expressed as 
\be 
q_1=\kappa \lambda_1^{1/2}, \quad 
q_2=\kappa \lambda_2^{1/2}. 
\en  
From Eqs.(3.7) and (3.8)  $\lambda_1$ and $\lambda_1$ 
 are the solutions of the  quadratic  equation,  
\be 
\lambda^2-(M^2+1 -\gamma_{\rm p}^2) \lambda +  M^2=0.  
\en 
Therefore,  $\lambda_1$ and $\lambda_1$  satisfy 
\be 
\lambda_1+\lambda_2= M^{2}+1-\gamma_{\rm p}^2, 
\quad 
\lambda_1\lambda_2= M^{2}. 
\en 
We also find   
$(\lambda_1-1)(\lambda_2-1)=\gamma_{\rm p}^2$, 
which will be used in the following calculations.  
Here the two parameters 
 $M$ and $\gamma_{\rm p}$  are   defined by   
\bea
&&M={1}/{\kappa \bar{\xi}}=[(\tau-\tau_c)/C]^{1/2}
/{\kappa} ,\\
&&\gamma_{\rm p}
= |g_a|/(4\pi C\ell_B)^{1/2} ,  
\ena  
where    $M \to 0$   as $\tau \to \tau_c$ and   
$\gamma_{\rm p}$ conveniently 
represents  the  solvation  asymmetry 
of the two ion species. That is, 
$\gamma_{\rm p}$ should be smaller than unity  
 for usual hydrophilic ion pairs    
 \cite{Nara,Current,Sadakane}.  
In  Appendix A, we shall see that 
our system is linearly unstable 
for $\gamma_{\rm p}>M+1$ 
against 
charge-density-wave formation, 
so we limit ourselves to 
the region $\gamma_{\rm p}<M+1$ 
in one-phase states.
In Fig.1, we display  $q_1/\kappa$ and $q_2/\kappa$ 
in the  $M$-$\gamma_{\rm p}$ plane. 

Some typical  cases 
are as follows. 
(i) As    $g_a \to 0$, 
 we have  
   $q_1 \cong \kappa$ 
and $q_2 \cong {\bar{\xi}}^{-1}$. 
(ii) Near the  criticality, we  have 
 $\kappa \bar{\xi }  \gg 1$ or $M\ll 1$. 
Furthermore, supposing  hydrophilic ions, 
we assume that  $\gamma_{\rm p}$ 
is  not close to unity and   
the inequality $1-\gamma_{\rm p}^2\gg M$ holds. We then find 
\be 
q_1 \cong \kappa ({1-\gamma_{\rm p}^2})^{1/2}, 
\quad q_2 \cong {\bar\xi}^{-1}({1-\gamma_{\rm p}^2})^{-1/2},
\en 
where $q_1 \gg q_2$ or $\lambda_1 \gg \lambda_2$. 
(iii) As shown in Fig.1, 
 $q_1$ and $q_2$  are complex conjugates 
in the region $|M-1|<\gamma_{\rm p}<M+1$. From Eq.(3.19) 
the real part $q_R=$Re$(q_1)$  
and the imaginary part  $q_I=$Im$(q_1)$ are calculated  as   
\bea 
&&
q_R =  [{(M+1)^2-
{\gamma_{\rm p}}^2}]^{1/2}\kappa/2,\nonumber\\ 
&&
q_I = [{{\gamma_{\rm p}}^2-(M-1)^2}]^{1/2}\kappa/2. 
\ena 
Oscillatory behavior appears for $q_I>q_R$ or for 
${\gamma_{\rm p}}^2>M^2+1$. 
As we approach the spinodal line or as 
$M \to {\gamma_{\rm p}}-1$ with ${\gamma_{\rm p}}>1$, 
$q_R$ becomes small as $(M-{\gamma_{\rm p}}+1)^{1/2}$, while  
  $q_I $ tends to nonvanishing 
$(\kappa /{\bar\xi})^{1/2}$.

\begin{figure}[t]
\begin{center}
\includegraphics[scale=0.45]{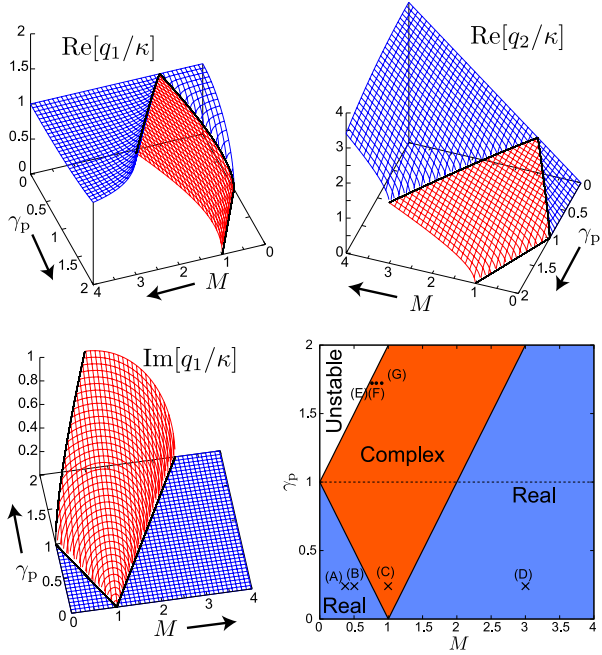}
\caption{(Color online) 
Real and imaginary parts of 
$q_1/\kappa$ and $q_2/\kappa$ 
in the plane of $\gamma_{\rm p}$ and 
$M=1/\kappa{\bar \xi}$ in the linear theory. 
Here  $q_1$ and $q_2$ are positive 
for $\gamma_{\rm p}<|M-1|$ (in blue) and are complex conjugates  
for $|M-1|<\gamma_{\rm p}<M+1$ (in red), while 
the system is unstable for $\gamma_{\rm p}>M+1$ (in white). 
These three regions are depicted in the right bottom panel, 
where the profiles and the interaction free energy 
will be shown at four $\times$
points (A), (B), (C) and  (D) 
 in Fig.2 and those at three $\bullet$ points (E), (F), and (G)  
 in Fig.3 in the linear theory.
}
\end{center}
\end{figure}

\subsection{Profiles around a single colloid particle}

In the presence of a single colloid  with radius $a$, 
we obtain the  fundamental profiles 
$U=U_0(r)$ and $\psi=\psi_0(r)$  induced by nonvanishing 
$\bar\gamma$ and $\bar\sigma$ from the boundary 
conditions (3.12) and (3.13). 
In the large system limit 
$V\gg a^3$, they  are expressed as linear combinations of two 
Yukawa functions, $e^{-q_1(r-a)}/r$ and $e^{-q_2(r-a)}/r$,  
where  $r$ is  the distance
from the colloid center.     As will be shown in Appendix B, 
they are of the forms, 
\bea 
&&\hspace{-1cm} U_0= 
\frac{g_aB_1 e^{-q_1(r-a)}}{(1-\lambda_1)(1+ q_1a)r}
-\frac{g_aB_2 e^{-q_2(r-a)}}{(1-\lambda_2)(1+ q_2a)r},
\nonumber\\
&&\hspace{-1cm} \psi_0=  
\frac{B_1e^{-q_1(r-a)}}{(1+ q_1a)r}
-\frac{B_2 e^{-q_2(r-a)}}{(1+ q_2a)r}. 
\ena 
The coefficients $B_1$ and $B_2$ are defined by  
\be
B_i  
= a^2[{\bar{\gamma}(1-\lambda_i)-\bar{\sigma}g_a}]/{C}({\lambda_1-\lambda_2}). 
\en   
In Appendix A, we will also 
express  the  correlation 
functions of the composition and the ion densities  
 as  linear combinations of these  
Yukawa functions.

We examine  some limiting cases. 
(i) As $g_a\to 0$, we find $\lambda_1 \to 1$, 
$\lambda_2 \to M^{2}$, and 
\be 
U_0 \cong  
{\bar Q}\frac{\ell_B e^{-\kappa (r-a)}}{(1+\kappa a)r},\quad 
\psi_0 \cong -{ {\bar\gamma}} 
\frac{a^2e^{-(r-a)/{\bar\xi}}}{C(1+ a/{\bar\xi})r},
\en 
where $U_0$ is the  
Debye-H$\ddot{\rm u}$ckel form \cite{Russel,LevinReview,Dej,Ov} 
with ${\bar Q}$ being  the average  charge of a  colloid particle, 
\be 
{\bar Q}= -4\pi a^2 {\bar \sigma}e . 
\en 
(ii) We  assume $|q_i |a\ll 1$ and $|q_i| (r-a)\ll 1$, 
which hold in  the limit of small $q_i$.  
Even for  $g_a\neq 0$, Eq.(3.25) yields 
the Coulombic behavior, 
\be 
U_0\cong {\bar Q}{ \ell_B} \frac{1}{r}, \quad  
\psi_0  \cong - ({\bar\gamma} { a^2}/{C})  \frac{1}{r},
\en 
which follow from   the  relations 
$B_1- B_2= - {\bar\gamma}a^2/C$ and  
$B_1/(1-\lambda_1) -B_2/(1-\lambda_2) = 
 -4\pi \ell_B {\bar\sigma}a^2/g_a$.  
(iii) Let us  approach  the instability line $   
M=\gamma_{\rm p}-1$ in Fig.1. In this case, both 
 $U_0$ and $\psi_0$ grow as 
Im$[(\lambda_1-\lambda_2)^{-1}] = -1/2q_Iq_R \propto 
(M-\gamma_{\rm p}+1)^{-1/2}$.  In this limit,  
the linear theory is valid only 
for  very   small 
$\bar \sigma$ and $\bar \gamma$.

\subsection{Interaction between two  colloid particles}

We suppose  two colloid particles of the same species 
with radius $a$. 
In  Appendix B, we will derive 
the  interaction free energy $F_{\rm int}$ 
from $\Delta\Omega$ in Eq.(3.17) as 
a linear combination of two Yukawa functions, 
 $e^{-q_1(d-2a)}/d$ and $e^{-q_2(d-2a)}/d$, 
 where $d$ is the  separation distance between the 
 two colloid centers 
 longer than   $2a$.  We  express it  as     
\be 
\frac{ F_{\rm int}}{T}= {4\pi a^4} 
\bigg[\frac{E_1e^{-q_1(d-2a)}}{(1+ q_1a)^2d}
- \frac{E_2e^{-q_2(d-2a)}}{(1+ q_2a)^2d}\bigg] ,
\en 
where the coefficients $E_1$ and $E_2$  are defined by  
\be 
E_i = \frac{[\bar{\gamma}(1-\lambda_i)-
\bar{\sigma}g_a]^2}{C(1-\lambda_i)(\lambda_1-\lambda_2)} .
\en 
Notice that $F_{\rm int}$   is independent of 
 the colloid dielectric constant 
$\ve_{\rm p}$. In Appendix B, 
we can see that  $\ve_{\rm p}$ appears in the third order 
contribution. 

Some limiting cases are as follows. 
(i) In the limit of weak solvation 
$g_1\to 0$ and  $g_2\to 0$, 
we have  $E_1 \to 4\pi \ell_B {\bar\sigma}^2$ 
and $E_2\to {\bar \gamma}^2/C$, leading to 
a decoupled expression, 
\bea 
{ F_{\rm int}}&=& 
F_{\rm DLVO}+ F_{\rm ad}  \nonumber\\
&&\hspace{-1cm} =  
\frac{{\bar Q}^2e^{-\kappa(d-2a)}}{\bar{\ve}(1+ \kappa a)^2d}
- 4\pi  
\frac{ Ta^4\bar{\gamma}^2 e^{-(d-2a)/{\xi}}}{C(1+ a/{\xi})^2d} .
\ena 
The first term  is the  DLVO   
interaction  $F_{\rm DLVO}$ in Eq.(1.2) 
\cite{LevinReview,Dej,Ov,Russel}. 
The second term  represents 
the adsorption-induced  attraction $F_{\rm ad}$  
for small $\bar\gamma$. 
 For neutral colloids,  we have 
 $\bar{\gamma}=\gamma$ 
and $h_1= -T\gamma$ to find the expression (1.5) 
with $A_{\rm ad}= 4\pi T { \gamma^2}/C$.     
Note that the linear theory is not applicable very 
 close to the criticality \cite{h1}, as stated below Eq.(1.5). 
(ii) 
When $|q_i |a\ll 1$ and $|q_i| (d-2a)\ll 1$, 
we use $E_1- E_2= {4\pi \ell_B}{\bar\sigma}^2- 
{{\bar\gamma}^2}/{C}$ to 
obtain   
 \be 
{F_{\rm int}}\cong 4\pi Ta^4({4\pi \ell_B}{\bar\sigma}^2- 
{{\bar\gamma}^2}/{C})/d . 
\en  
(iii) Near the criticality and for hydrophilic ions, 
we may assume  $M\ll 1$ and  
 $1-\gamma_{\rm p}^2\gg M$, where    
$q_1$ and $q_2$ 
are given by  Eq.(3.23). Then ${F_{\rm int}}$ in Eq.(3.30) 
takes the  same form as the decoupled expression (3.32):   
\be 
{ F_{\rm int}}= 
\frac{{\bar Q}_e^2e^{-q_1(d-2a)}}{\bar{\ve}(1+ q_1 a)^2d}
- 4\pi  
\frac{ T a^4\bar{\gamma}_e^2 e^{-q_2(d-2a)}}{C(1+ q_2a)^2d} .
\en 
where ${\bar Q}$ and $\bar{\gamma}$ 
in Eq.(3.32) have been replaced by 
 ${\bar Q}_e$ and $\bar{\gamma}_e$ defined by 
\bea 
&&{\bar Q}_e= {\bar Q}(1-g_a {\bar\gamma}/
4\pi C\ell_B{\bar\sigma})/\sqrt{1-\gamma_{\rm p}^2},\nonumber\\
&& 
\bar{\gamma}_e= ({\bar \gamma}- {\bar\sigma}g_a) 
/\sqrt{1-\gamma_{\rm p}^2}.
\ena 
(v)  When  $q_1$ and $q_2$  are complex conjugates 
in the region $|M-1|<\gamma_{\rm p}<M+1$ in Fig.1, 
Eq.(3.30) gives  
\bea 
&&\hspace{-1cm}
{ F_{\rm int}} =
8\pi T a^4 \exp[{-q_R (d-2a)}]\frac{1}{d} \nonumber\\
&&\hspace{-1cm}\times\{ J \cos [q_I (d-2a)] 
+K \sin [q_I (d-2a)] \},
\ena 
where the coefficients 
$J$ and $K$ are the real and imaginary parts 
of $E_1/(1+q_1a)^2$. 
On approaching the spinodal line $\gamma_{\rm p}= M+1$, 
$K$ grows as $ 1/q_R$ but $J$ remains finite.  

%
For neutral colloids, 
we compare the van der Waals  interaction 
$F_{\rm vdw}(d)$ in Eq.(1.2) 
and the adsorption-induced 
interaction $F_{\rm ad}(d)$ in Eq.(1.5) or  (3.32) 
at the closest separation $d-2a=a_0$, 
where $\bar{\gamma}=\gamma$ and $\bar{\xi}=\xi$.  
 If  $C\sim a_0^{-1}$  \cite{comment-gra}, 
we estimate their ratio as
\bea 
\frac{F_{\rm ad}}{F_{\rm vdw}}  
&&\sim  10^{2}(a_0a\gamma)^{2}T/{A_H} 
 \quad (a\gg \xi) \nonumber\\
&&\sim  10^{2} (a_0 \xi \gamma)^{2}T/{A_H}   
 \quad (a\ll \xi). 
 \ena 
Even at the closest separation, 
the van der Waals attraction  is negligible 
when $A_H/T$ is 
smaller than $(10a_0a\gamma)^2$ for $a\gg \xi$ 
and  than $(10a_0\xi\gamma)^2$ for $a\ll \xi$.

\subsection{Plotting analytic 
results in the linear theory}
\begin{figure}[t]
\begin{center}
\includegraphics[scale=0.45]{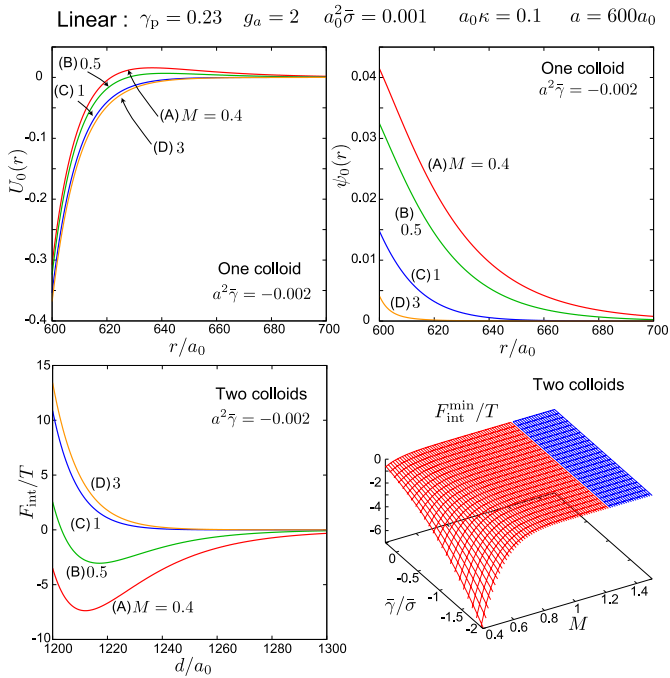}
\caption{(Color online)  Results of the linear theory 
 with $a=600a_0$ for hydrophilic  ions 
 with  $\gamma_{\rm p}=0.23$, $g_a=2$, 
 $\bar{\sigma}=0.001 a_0^{-2}$, 
 and $\kappa=0.1a_0^{-1}$, where $M$ is 
 (A) 0.4, (B) 0.5, (C) 1, and (D) 3 
(see Fig.1 for their locations in the $M$-$\gamma_{\rm p}$ 
plane). Top: Normalized potential $U_0(r)$  (left) and 
composition deviation $\psi_0(r)$  (right)  in Eq.(3.25) 
vs $r/a_0$ around  a  colloid. 
Left bottom: Interaction free energy $F_{\rm int}(d)$ 
between two colloids in Eq.(3.30)   vs $d/a_0$.  
 Right bottom: Minimum 
 of $F_{\rm int}$,  denoted by $F_{\rm int}^{\rm min}$, in the    
 $M$-$\bar{\gamma}/\bar{\sigma}$ plane with the  common 
 $\bar{\sigma}$,  $\kappa$,  and $a$. 
 It  is zero (attained at infinity) 
 in the right (for $M\gs 1.3)$  
 and is negative in the left (for smaller $M$). 
}
\end{center}
\end{figure}

\begin{figure}[t]
\begin{center}
\includegraphics[scale=0.45]{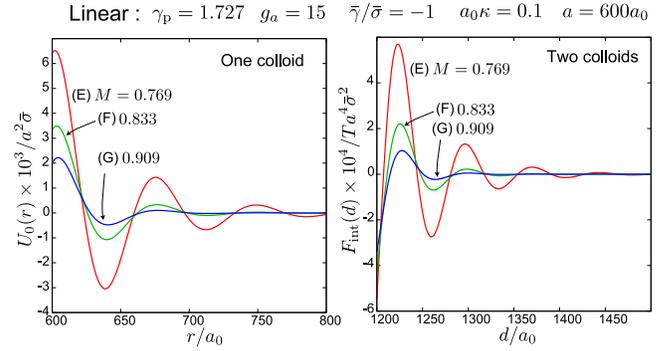}
\caption{(Color online) Results of the linear theory 
 with $a=600a_0$ for 
antagonistic ions  with   $\gamma_{\rm p}=1.727$, $g_a=15$, 
$\bar{\gamma}/\bar{\sigma}=-1$,  
 and $\kappa=0.1a_0^{-1}$, where  
 $M$ is (E) 0.769, (F) 0.833, 
 and (G) 0.909 (see Fig.1 for their locations). 
 Left: 
Normalized potential $U_0(r)/a^2{\bar\sigma}$  
vs $r/a_0$ around  a colloid.  
Right: Normalized 
interaction free energy $F_{\rm int}(d)/a^4{\bar\sigma}^2$ 
between two colloids from  Eq.(3.30)   vs $d/a_0$.  
}
\end{center}
\end{figure}

On the basis of the analytic expressions (3.25) and (3.30),  
 we plot $U_0(r)$ and $\psi_0(r)$ 
around a single colloid  
and $F_{\rm int}$ between two colloids   
for various $M$.  Assuming 
a large radius $a=600a_0$, 
we set $\ell_B=3a_0$   
and  $\kappa=0.1a_0^{-1}$, where $a_0=v_0^{1/3}$.  
Then $n_0=1.33 \times 10^{-4}v_0^{-1}$.  

In  Fig.2,  with   hydrophilic ions, we set 
 $\gamma_{\rm p}=0.23$, $g_a=2$,  
 $\bar{\sigma}=0.001 a_0^{-2}$, and 
 $\bar{\gamma}/\bar{\sigma}=-2$. 
We can see that 
 $\psi_0$   increases  considerably with 
 decreasing  $M$, while $U_0$ is rather insensitive to $M$.  
Remarkably,  the curves of $F_{\rm int}$ vs $d/a$ 
exhibit  a negative minimum at an intermediate 
$d$  for small $M$. 
With  these selected parameters, 
 the DLVO interaction 
 $F_{\rm DLVO}$  is  equal to 
$1.061 \times  10^{-4}T a^4{\bar\sigma}^2=13.75 T$ at $d=2a_0$.

In  Fig.3,  with  antagonistic  ions \cite{Nara,Current,Sadakane},  
we display 
$U_0(r)/a^2{\bar\sigma}$  
and $F_{\rm int}(d)/Ta^4{\bar\sigma}^2$
by setting $\gamma_{\rm p}=1.727$, $g_a=15$, 
and $\bar{\gamma}/\bar{\sigma}=-1$. 
 As   $M \to M_c=0.727$, $U_0(r)$ and $F_{\rm int}(d)$ grow 
 as stated below Eqs.(3.24) and (3.36). Oscillatory 
 relaxation is conspicuous close to the 
 instability. Here,  the linear conditions,  $|U_0| \ll 1$ 
 and $|\psi_0| \ll 1$, are  ensured only by 
 very small $\bar\sigma$ and $\bar\gamma$. Otherwise, 
   the nonlinear theory is required.  
In Fig.3,  we thus divide $U_0$ and $F_{\rm int}$ by 
$a^2{\bar\sigma}$ and $a^4{\bar\sigma}^2$, respectively. 
 
\section{Precipitation due to 
 highly selective solvation}

In a binary mixture in one-phase states, 
a  highly  selective solute can induce 
 precipitation of domains 
rich in  the preferred  component \cite{Okamoto,Current}.  
The solute can be  either a hydrophilic salt 
(such as NaCl)  or 
  a neutral hydrophobic solute. 
The equilibrium size of a precipitated domain  
depends on the system size $V$ (as can be known from 
Eq.(4.8) below).  

In the next section, 
we  will show numerically that 
precipitation can occur   on the colloid  surface.  
Therefore, in this section,  we   summarize 
and extend our previous theory \cite{Okamoto} 
to understand precipitation on the colloid surface.  
We  suppose an experiment 
of  very dilute colloid suspension; then,  
  the  volume $V$ assigned to each colloid particle 
is   the inverse  droplet density $n_D^{-1}$.  For 
high  colloid concentrations,  
colloids should interact collectively 
due to the growth  of wetting layers 
to form a floccuated phase \cite{Beysens,Guo,Kaler}.

\subsection{Bulk precipitation}
\setcounter{equation}{0}

For   hydrophilic cations and anions,   the 
phase behavior of precipitation   is little affected 
by charge density variations, because they 
are significant   only at 
interfaces, colloid surfaces,  or a  
container (see the right bottom panel of Fig.10 as an  
example).  Thus, neglecting the electrostatic interaction,  
we may use results of 
a three component system  \cite{Okamoto} by setting 
\bea  
&& n_1=n_2=n/2, \\
&&  g=(g_1+g_2)/2 .   
\ena 
We assume    the strong solvation condition $g\gg 1$. 
 In the precipitated  phase, called  $\alpha$, 
the water composition $\phi_\alpha$   is close to unity 
and the solute density $n_\alpha$
 is much larger than the average ${\bar n}=2n_0$ by 
  the  factor $e^{g(1-{\bar\phi})}$ \cite{largefactor}.  
 The precipitation  effect is significant   
 for not small $g(1-{\bar\phi})$.

Let us  decrease  $\bar n$ in the presence of 
precipitated domains. Then the  
 volume fraction 
$\gamma_\alpha$ of  phase $\alpha$ decreases 
and eventually tends to zero 
as 
\be 
\gamma_\alpha 
=({\bar n} / n_{\rm p}-1)e^{-g (1-\bar{\phi})} ,
\en 
where  $n_{\rm p}$ is a minimum solute density for precipitation. 
Its asymptotic expression for $g\gg 1$ is given by 
\be 
n_{\rm p}= e^{-g (1-\bar{\phi})}G({\bar \phi},\chi)/ T. 
\en 
The function $G({\bar \phi},\chi)$ is a positive quantity 
  defined   by   
\bea 
G({\bar \phi},\chi) &=& -f(\bar \phi) - 
(1-\bar \phi)f'(\bar \phi)\nonumber\\
&=& - (T/v_0) [\ln\bar  \phi+ \chi(1-\bar \phi)^2], 
\ena 
where   the second line follows   from Eq.(2.4). 
Alternatively, we may  decrease $\chi$ 
at fixed $\bar \phi$ and ${\bar n}$ 
in the presence of precipitated domains. 
The volume fraction $\gamma_\alpha$ 
tends to zero  as  $\chi $  approaches 
a lower bound  $\chi_{\rm p}=
\chi_{\rm p}(\bar{\phi}, {\bar n})$. From Eq.(4.4),  
$\chi_{\rm p}$  satisfies 
$
{\bar n} = e^{-g (1-\bar{\phi})}G({\bar \phi},\chi_{\rm p})/ T.
$   
Use of the second line of Eq.(4.5)  gives   
\be
\chi_{\rm p}=  -[ \ln {\bar\phi}+ e^{g (1-\bar{\phi})} v_0{\bar n}
]/(1-{\bar\phi})^2. 
\en 
It  also follows  the relation, 
\be 
\chi-\chi_{\rm p}\cong 
v_0 e^{g(1-{\bar \phi})}({\bar n}-n_{\rm p})/(1-{\bar\phi})^2.
\en 
Here  $\bar n$ appears in the combination 
 $e^{g (1-\bar{\phi})}{\bar n}$.

\begin{figure}[t]
\begin{center}
\includegraphics[scale=0.45]{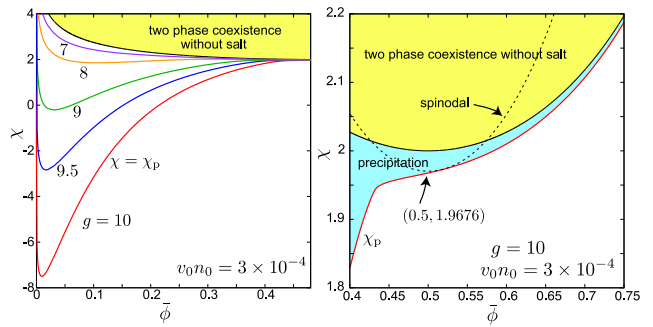}
\caption{(Color online) 
Minimum interaction parameter $\chi_{\rm p}$ 
vs $\bar\phi$ at ${n}_0= 3\times 10^{-4}v_0^{-1}$ 
in the range $0<{\bar\phi}<0.47$  
for $g=7, 8, 8.5, 9$, 9.5, and 10 
from above (left) 
and in the range $0.4<{\bar\phi}<0.75$ 
for $g=10$ (right). 
Two-phase coexistence region (in yellow)  without solute is 
in the upper part. Spinodal curve $\tau=\tau_c$ 
is also written by dotted line (right).  
}
\end{center}
\end{figure} 

In Fig.4, we display  
 $\chi_{\rm p}$ vs $\bar\phi$ 
  at ${n}_0= 3\times 10^{-4}v_0^{-1}$. 
The left panel gives the curves   
for $g=7, 8, 8.5, 9$, 9.5, and 10  
in the range $0<{\bar\phi}<0.47$. 
With increasing  $g$, 
 the precipitation branch 
is suddenly detached downward from the solvent 
coexistence curve. For $g \gg 1$, they 
are in excellent agreement with the 
asymptotic formula  (4.6) for $\chi_{\rm p}$. 
(The asymptotic formula (4.4) for $n_{\rm p}$ 
is also a good approximation for 
$g \gg 1$ \cite{Okamoto,Current}.)
The right panel gives the curve for 
$g=10$ at the same $n_0$ 
in the range $0.4<{\bar\phi}<0.75$.  
The curve is only slightly outside the coexistence 
curve for ${\bar\phi}\gs 0.6$ 
and  touches 
the spinodal curve $\tau=\tau_c$ 
at $(\chi, {\bar\phi})= (0.5187, 1.9728)$ \cite{touch}. 
We obtained these  curves 
numerically from  minimization of the bulk free energy 
at $\av{\phi}={\bar \phi}$ and $\av{n}={\bar n}$,  
 neglecting  the surface free energy. 
 Thus  these curves 
are those for two-phase coexistence 
with a planar interface separating the two phases.

\subsection{Surface tension effect on a precipitated droplet}

The surface tension $\sigma_s$ between a 
 precipitated droplet and the surrounding solute-poor 
 region is well-defined, though the precipitation 
 is a nonlinear effect of a highly selective solute.  
  It  is of order $\sigma_s \sim T/a_0^2$ 
  far from the solvent criticality, being 
independent of the solute density.  
We here  examine the surface tension 
effect on the droplet stability.
 
We consider   a single spherical droplet 
of phase $\alpha$ with radius $R$ in a large volume $V$.
 The droplet    volume fraction is 
then  $\gamma_\alpha =v/V$, where $v= 4\pi R^3/3$. 
For $\bar n>n_{\rm p}$ and at very small  volume fraction 
$ \gamma_\alpha \ll e^{-g(1-\bar\phi)}$, 
the droplet free energy   
is expressed as \cite{comment1,Binder} 
\bea 
&&\hspace{-5mm}{\Delta F} =
 4\pi \sigma_s {R^2}-T(\bar n-n_{\rm p}) e^{g(1-\bar\phi)}v 
 +  \frac{T{\bar n}}{2V}e^{2g(1-\bar\phi)}v^2
 \nonumber\\ 
 &&= 4\pi \sigma_s(R^2 -2R^3/3R_c + R^6/3R_{\rm m}^4). 
\ena 
In the first line, the first two terms constitute  the 
standard  droplet free energy in 
the nucleation theory and the third  term ($\propto R^6/V$)
arises from the finite size effect \cite{Onukibook}. 
In the second line, we introduce 
 a critical radius $R_c$ and a minimum radius 
 $R_{\rm m}$ by 
\bea
&&
R_c= 2\sigma_s e^{-g(1-\bar\phi)} /T(\bar n-n_{\rm p}) \nonumber\\ 
&&~~~ = 2v_0 \sigma_s /[T(1-{\bar\phi})^2(\chi-\chi_{\rm p})],\\
&& R_{\rm m} =  (3\sigma_s V/2\pi T{\bar n})^{1/4}
e^{-g (1-\bar\phi)/2} .
\ena 
where  Eq.(4.7) is used in the second line of Eq.(4.9), 
$R_c$ grows as $\chi\to \chi_{\rm p}$, and $R_{\rm m}$ 
decreases with decreasing $\bar\phi$. 
We estimate $R_{\rm m} \sim
(Va_0/ v_0 {\bar n})^{1/4}  
e^{-g(1-{\bar\phi})/2}$ 
far from the solvent criticality.
Minimization of  the second line of Eq.(4.8) 
 yields  
\be 
(R/R_{\rm m})^4= R/R_c-1>0
\en 
 for  the equilibrium  radius $R$. 
 We require $\Delta F/4\pi \sigma_s 
 R^2= (2 -R/R_c)/3<0$ 
 to find  $R>2R_c$. 
 Hence $R>R_{\rm m}$ from Eq.(4.11) and 
 $R_{\rm m}$  is the minimum radius of equilibrium  droplets  
 in a finite system \cite{comment1}.  For $R_{\rm m} \gg R_c$, 
 droplets with radii 
much larger than $R_{\rm m}$ can appear 
as 
\be 
R^3\cong R_{\rm m}^4/R_c,  
\en  
where  the right hand side is proportional to $V$ 
and is independent of $\sigma_s$ 
in agreement with Eq.(4.3).

 It also follows the  condition  
 $ R_{\rm m}>2R_c$ from Eq.(4.11) \cite{comment2},  
 leading  to  lower bounds 
 of ${\bar n}-n_{\rm p}$ and $\chi-\chi_{\rm p}$ as  
\bea 
&&{\bar n}-n_{\rm p}>
4\sigma_se^{-g(1-\bar\phi)} /TR_{\rm m} ,\\
&&\chi-\chi_{\rm p} 
>4v_0\sigma_s /TR_{\rm m}(1-\bar\phi)^2,
\ena  
for the formation of a  droplet. 
Thus the bulk precipitation curves ${\bar n}=n_{\rm p}$ 
 and ${\chi}=\chi_{\rm p}$   
 are shifted upward   
 by amounts proportional to $\sigma_s/R_{\rm m} 
\propto \sigma_s^{3/4}V^{-1/4}$ for droplets 
with surface tension. 

As an example, let us set $V=4\times 10^9 \pi a_0^3/3$ 
with  $C= 2a_0^{-1}$, $g=10$, 
 $\bar{n}=6\times 10^{-4}v_0^{-1}$, 
 and $\bar\phi=0.41$. The 
  surface tension is then $\sigma_s= 0.06 T/a_0^2$ as  
 $\chi \to \chi_{\rm p}=1.871$ \cite{Okamoto}.  
 Thus  we obtain $n_{\rm p}=6\times 10^{-4}$ and 
 $R_{\rm m}=35.1a_0$. The  right hand side of Eq.(4.13) 
is  $1.9\times 10^{-5}$, while that 
of Eq.(4.14) is $0.02$.


\subsection{Wetting layer formation on a colloid surface}

A  completely wetting   layer can appear 
on a colloid surface   above a   precipitation 
 curve for the hydrophilic  case 
 $\gamma<0$ or for the hydrophobic case $\gamma>0$ 
 under the condition (2.29). 
 In our numerical analysis, 
 the precipitation curve 
for a colloid is   only 
slightly shifted upward from the bulk  
curve $\chi=\chi_{\rm p}$ in the $\chi$-$\bar\phi$ plane. 
We also recognize that  the precipitation curve 
is nearly independent of the surface parameters 
$\gamma$ and $\sigma_0$ (see Figs.6 and 8). 

We suppose that  a colloid with radius $a$  
is  completely  wetted by   a 
 spherically symmetric layer with thickness $R-a$. 
The volume fraction of phase $\alpha$ 
is $\gamma_\alpha= 4\pi (R^3-a^3)/3V$ in a volume $V$. 
For   $R-a \gg a_0$,  we  may treat 
 the surface free energy between the colloid surface and 
 the wetting layer  as  a constant. 
As a generalization of  Eq.(4.8),  $R$ is determined by 
minimization of a  free energy contribution 
$F_{\rm wet}$. In terms of $R_c$ in Eq.(4.9) 
and $R_{\rm m}$ in Eq.(4.10), it is expressed as  
\bea 
&&\hspace{-4mm} \frac{ F_{\rm wet}}{ 4\pi \sigma_s}= R^2-a^2
- \frac{2}{3R_c}(R^3-a^3)+ \frac{1}{3R_{\rm m}^4}(R^3-a^3)^2 
\nonumber\\
&&\hspace{0.4cm}= 
{  R_{\rm m}^2}\bigg[  
 (q+ s^3)^{2/3}-s^{2} 
- \frac{2R_{\rm m}}{3R_c}q + \frac{q^2}{3} \bigg]. 
\ena 
The first line tends to  Eq.(4.8) as  $a\to 0$. 
In the second line,  $q$ and $s$ are defined by  
\bea
q &=&  (R^3-a^3)/R_{\rm m}^3,\\
s &=&  a/R_{\rm m}. 
\ena 
We treat $q$ as an order parameter. 
In Eq.(4.15),  the selective solvation 
is accounted for in   $R_c$  
and $R_{\rm m}$,  but the electrostatic interaction 
is neglected. See Fig.18 in Ref.\cite{Okamoto} for 
 $q$ in the  $s$-$R_{\rm m}/R_c$ plane (where 
 $s$ in Eq.(4.17) is written as $D$). 
In the thin layer limit  $R^3/a^3 -1\ll 1$ 
or for $q \ll s^3$,  $ F_{\rm wet}$ is 
expanded  up to the third order with respect to $q$ as 
\be 
\frac{ F_{\rm wet}}{ 4\pi \sigma_s R_{\rm m}^2}=
{(1-\frac{a}{R_c})} \frac{2q}{3s}  
+ ({3} -  \frac{1}{s^{4}})\frac{q^2}{9} + \frac{4q^3}{81 s^{7}}.
\en 
Here the coefficients of the first two 
 terms can vanish at $a/R_c=1$  and $s=3^{-1/4}$, 
 where we  predict   tricritical behavior 
 with varying $\chi$ or $ R_c$. 
 That is, $q$ changes 
 continuously or discontinuously  
depending on 
whether $a>3^{-1/4}R_{\rm m}$ or $a<3^{-1/4}R_{\rm m}$.

For  $a>  
3^{-1/4}R_{\rm m},
$  
$q$ becomes nonvanishing 
for $1-a/R_c<0$ or for 
$R_c^{-1}>a^{-1}$  continuously 
as a second-order phase transition. 
From Eq.(4.9) this 
condition yields  a lower bound of  
$ \chi-\chi_{\rm p}$  in the form, 
\be 
 \chi-\chi_{\rm p} 
>2v_0\sigma_s /T a (1-\bar\phi)^2,  
\en   
which also follows if  $R_{\rm m}$ is  replaced  by $2a$ 
in  Eq.(4.14). With further increasing $a$ 
much above $R_{\rm m}$ or for    
$s\gg 1$, the first two terms in 
Eq.(4.18) gives 
$ q \cong  (a/R_c-1)/s$ or  
\be 
R^3/a^3-1 \cong  (R_{\rm m}/a)^4( a/R_c-1),  
\en 
which holds for 
$a/R_c-1\ll (a/R_{\rm m})^4$ 
because we have 
assumed  $q \ll s^3$  in setting up  Eq.(4.18). 
The situation $a\gg R_{\rm m}$ 
can well happen  in  real systems particularly for 
relatively small $\bar\phi$ (where $R_{\rm m}$ 
is decreased). In such  cases, 
the factor $(R_{\rm m}/a)^4$ in Eq.(4.20) is  small and  
the wetting layer  thickens  slowly 
with increasing $\chi$. 
On the other hand, in the thick 
layer limit  $R\gg a$, $R$ is given by  Eq.(4.12).

For       
$a<3^{-1/4}R_{\rm m},
$  
 $q$ becomes nonvanishing   discontinuously 
as a first-order phase transition  
when  $1-a/R_c $ is smaller than 
a small positive constant 
 ($\propto (s-3^{-1/4})^2$ 
near the tricriticality). As a result, 
the lower bound of $ \chi-\chi_{\rm p}$ 
for precipitation 
is slightly  smaller than the right hand side of Eq.(4.19).   

In the above theory, 
we have examined  the transition  between 
weakly adsorbed states and  completely wetted 
states. In the next section, however, 
we shall see 
that a hydrophobic  surface can be partially wetted 
by a nonspherical 
water-rich layer 
at relatively small $\bar\phi$ 
and for not very large $\gamma$ (see  Figs.16 and 17).

\section{Numerical results }
\setcounter{equation}{0}

In this section,  we present numerical results  
on the basis of our nonlinear theory in Sec.II with 
 a single colloid or two colloids placed at the center of 
a large cell. The correlation length 
$\bar\xi$ in Eq.(3.11) increases up to $14.1 a_0$ 
at $\chi= 1.965$  in Fig.12, but it 
is only   a few times 
longer than $a_0$ in the other examples far from the solvent 
criticality.

\subsection{Parameter values selected}
 We set 
$C=2a_0^{-1}$, $\Delta_0=18$,  $\Delta_1=12$, 
$\ell_B =e^2/\ve_0 k_BT=3a_0$, and $\ve_1=\ve_0$. 
We use large  $\Delta_1 $, 
so  the degree of ionization 
$\alpha$ sensitively  
 depends  on the surface composition from Eq.(2.25). 
Several values are 
assigned to the density of ionizable groups.  
surface with $\gamma>0$. Because of the numerical convenience, 
the colloid radius $a$ is assumed  to be rather small as 
\be 
a=15a_0.
\en 
We also performed simulations with  $a=25a_0$ 
to obtain essentially the same results, though 
the corresponding figures are not shown.

 In  Subsecs.VB-D, 
we treat hydrophilic ions  with 
$g_1=11$ and  $g_2=9$ at  $v_0{n}_0=3\times 10^{-4}$ 
in Figs.5-17.   
Recall that we  have introduced 
the Debye wavenumber  $\kappa$ in Eq.(3.5) and 
the asymmetry parameter $\gamma_{\rm p}$ in Eq.(3.22) 
as functions of the average composition $\bar\phi$. 
Here,  we have  
 $\kappa=0.127a_0^{-1}$ and $\gamma_{\rm p}= 0.135$ 
 for  $\bar\phi=0.4$.  
 In  Subsec.VE,  we treat  antagonistic 
ion pairs with  $g_1=-g_2=13$  at 
$v_0{n}_0=  10^{-3}$ in  Figs.18 and 19, where   
we  have  $\kappa=0.233a_0^{-1}$ and 
  $\gamma_{\rm p}= 1.76$ for $\bar\phi=0.4$.

 In  the case of a single colloid, 
our cell   is a large sphere 
with radius  $10^3a_0$ 
 in the spherically symmetric  geometry. 
 In  the case of two  colloids,    it is a cylinder   
with radius $256a_0$ and  
 height  $512a_0$  in the axisymmetric  geometry. 
Since $a=15a_0$, the colloid volume fraction 
is  $3.375\times 10^{-6}$ for 
 a single colloid and is  $1.341 \times 10^{-4}$ for  
two  colloids.  For hydrophilic ions 
and  at $\bar\phi=0.41$,  
$R_{\rm m}$ in Eq.(4.10) is  $35.1a_0$  in the  
single colloid case \cite{largefactor} 
and $13.0a_0$ in the two  colloid case.   
In agreement with the discussion  in Subsec.IVC,    
the layer formation due to 
precipitation is discontinuous 
for  the  single colloid case 
(where $a<3^{-1/4}R_{\rm m}$) 
but  is continuous for 
the two colloid case 
(where $a>3^{-1/4}R_{\rm m}$).

\subsection{Comparison of  the linear and nonlinear theories}

\begin{figure}[t]
\begin{center}
\includegraphics[scale=0.42]{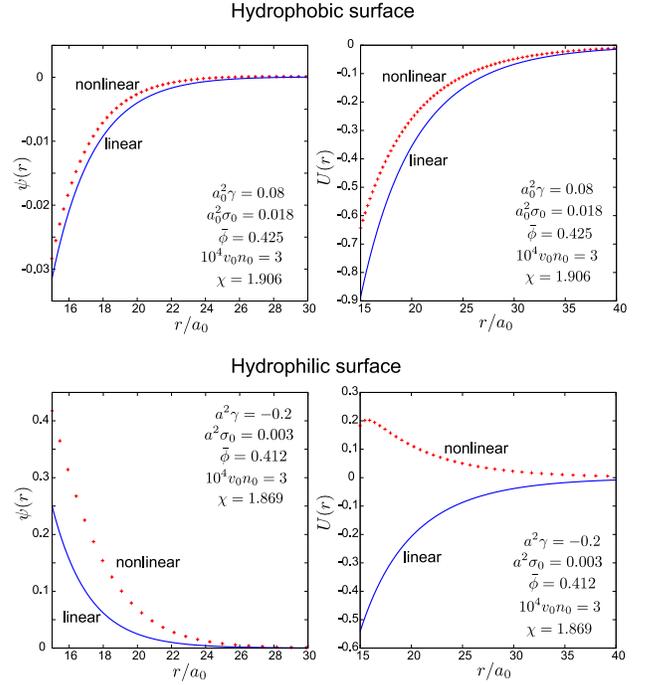}
\caption{(Color online) 
 $\psi(r)=\phi(r)-{\bar\phi}$ and $U(r)$ 
vs $r/a_0$ around a colloid 
from  the linear  and  nonlinear theories  
below the precipitation line  at 
$v_0{n}_0=3\times 10^{-4}$. 
Upper plates: Those 
for a hydrophobic  surface with $\gamma=
0.08a_0^{-2}$ 
at $\sigma_0=0.018a_0^{-2}$, where 
  $|\psi|$  at the surface is  smaller than $1/g_i$ 
and $|U|$ is at most of order unity so that 
 the linear results 
 fairly agree with the nonlinear results.  
Lower plates: Those   for a hydrophilic  surface with 
 $\gamma=0.2 a_0^{-2}$ at small 
 $\sigma_0=0.003a_0^{-2}$. From 
$\psi\sim 0.3$ as $r \to a$, the cations are considerably  accumulated 
  near the surface and $U$ becomes  positive 
 in the nonlinear theory, while $U$ remains  negative 
 in the linear theory.  
 } 
\end{center}
\end{figure}

We compare the profiles 
of $\psi(r)=\phi(r)-{\bar\phi}$ and 
$U(r)=e\Phi(r)/T$ from the 
linear theory in Sec.III and those from the 
nonlinear theory in Sec.II below the bulk precipitation curve 
$\chi<\chi_{\rm p}$.  The linear theory 
 is based on the assumptions (3.2) and (3.3) 
 and is thus valid  only 
 for $|\psi| \ls 1/|g_i|$ and $|U|\ls 1$ near 
the surface.

In the upper plates of Fig.5, we suppose  
 hydrophilic ions and 
a hydrophobic  surface with $\gamma=0.08a_0^{-2}$.
The other parameters are  $\chi=1.908$, 
 $\bar\phi=0.425$,   and ${\sigma}_0=0.018 a_0^{-2}$.  
Here  $\alpha= 0.25$ from  the nonlinear calculation,  
leading to  ${\bar\sigma} = 0.0045$ 
 and ${\bar\gamma} = 0.026$, which were then 
 used  in   the linear calculation.  
In this case, $g_i \phi$ and  $U$ 
are both  decreased near the  surface 
and their amplitudes are not large compared to unity. 
Thus the linear results are in fair agreement 
with the  nonlinear results.

In the lower  plates of Fig.5, we suppose  
  hydrophilic ions and  
a hydrophilic  surface with  $\gamma=-0.2a_0^{-2}$.  
 The other parameters are  $\chi=1.869$, 
 $\bar\phi=0.412$, and ${\sigma}_0= 0.003 a_0^{-2}$. 
We have  
$\alpha= 0.992$ in the nonlinear calculation, which gives 
  ${\bar\sigma} =0.00297a_0^{-2}$ 
 and ${\bar\gamma}=-0.236a_0^{-2}$.  
  In this case, $\phi$ is a few times larger than 
 $1/g_i\sim 0.1$ near the  surface,   
 leading to considerable  ion accumulation, 
 which cannot be accounted for in the linear theory. 
 Furthermore, the cations are more enriched 
  than the anions near the surface, 
leading to positive $U$ in the nonlinear theory, 
while $U$ remains  negative in the linear theory.  

Comparison of 
the linear and nonlinear theories 
will also be made 
in the presence of two colloids at $\bar{\phi}=0.5$. 
See explanations  of Fig.12 below.  

\subsection{Prewetting and precipitation on a colloid} 

With hydrophilic ions,  
there appear   two transition  lines 
of  prewetting and  precipitation    for each colloid 
in the ${\bar\phi}$-$\chi$ plane. They are located   
far below the solvent  coexistence curve 
for strong selective solvation. 
The  prewetting  line  
sensitively depends  on $\gamma$ and $\sigma_0$ (as 
will be  seen in Figs.6 and 8 below). 
It     starts   from a point on 
the precipitation  line   ending  
at a surface critical point, 
across which there are discontinuities 
in the physical quantities.
The transition across the precipitation line 
for a colloid is  first-order 
for  the present parameters.  
 We further  confirmed  that the precipitation line  
 exhibits  no appreciable 
dependence on $\gamma$ and $\sigma_0$  
and approaches   the bulk one  $\chi=\chi_{\rm p}$   
with increasing  $a$ for  $a<3^{-1/4}R_{\rm m}$. 
If $a>3^{-1/4}R_{\rm m}$, 
the precipitation transition becomes  continuous 
and it is difficult to precisely determine 
the location of the precipitation line.

\subsubsection{Hydrophilic surface with $\gamma=-0.2a_0^{-2}$}

In our previous work\cite{Okamoto}, we 
examined one example of  a hydrophilic  colloid  
with $\gamma=-0.2a_0^{-2}$ in the presence of  
 hydrophilic ions, where   
precipitation and prewetting are    
first induced on the colloid surface 
before  in the bulk region. 
Again with $\gamma=-0.2a_0^{-2}$,  
 the left panel of Fig.6 
 displays  three examples of the 
 prewetting line  
corresponding  to $a^2\sigma_0=0$, 
0.003, and 0.005, around  which $\alpha$ is nearly 
equal to unity and $\sigma\cong \sigma_0$. 
A first-order precipitation line is 
also shown (broken  line), which is independent of $\sigma_0$. 
The right panel of Fig.6 shows 
 the preferential adsorption $\Gamma$  
around a colloid  as a function of $\chi$ 
for several $\bar \phi$ at $a^2\sigma_0=003$. 
Since $\phi (r)$ tends to a constant  
$\phi(L)$  far from the 
colloid, $\Gamma$ is calculated from    
\be 
\Gamma=  \int'   d{\bi r} [\phi({\bi r}) -{\phi}(L)], 
\en 
where the integration is in the colloid exterior. 
With  precipitation, 
we have $\Gamma \sim 
4\pi (1-{\bar\phi})(R^3-a^3)/3$ for a single colloid.

In Fig.7, profiles of $\phi(r)$ 
and the  total ion density $n_1(r)+n_2(r)$ are displayed 
at four points (A)-(D) marked in Fig.6, where  
 $\gamma=-0.2a_0^{-2}$,  $\bar\phi=0.412$,  
and $\sigma_0 =0.003 a_0^{-2}$. 
 We can see  a weakly discontinuous 
 prewetting transition between 
(A) and (B) and a strongly discontinuous 
precipitation transition   between 
(C) and (D).  
The value of $\chi$ at this  precipitation transition 
  is 1.8945,  
  while  it is  $1.8926$ for $d=25a_0$ 
  with  the   other parameters 
  held at the same values (not shown here). 
These values of $\chi$  are  
slightly larger than $\chi_{\rm p}=1.8795$ at $\bar{\phi}=0.412$.

\begin{figure}[t]
\begin{center}
\includegraphics[scale=0.46]{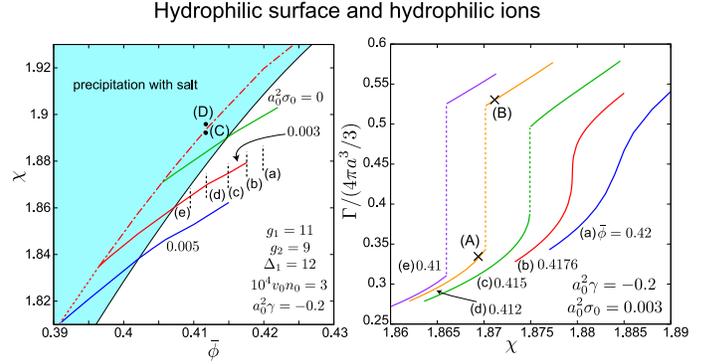}
\caption{(Color online) 
Prewetting and precipitation transitions on a 
 hydrophilic colloid surface with  $a=15a_0$ and 
 $\gamma=-0.2a_0^{-2}$. 
Left:Colloidal precipitation line (red broken 
line) and prewetting lines  for   $a_0^2\sigma_0=0$, 
0.003, and 0.005, around which  $\alpha\cong 1$ here. 
 Right: Normalized adsorption $\Gamma/(4\pi a^3/3)$ 
 vs  $\chi$  at $a_0^2\sigma_0=0.003$ 
for (a)  $\bar\phi=0.42$, (b) 0.4176, (c) 0.415, (d) 0.412, 
and (e) 0.41. No phase transition appears on  path (a), 
   the prewetting critical point is on path (b),  and 
 the transition  is  discontinuous  for paths (c), (d), and (e). 
}
\end{center}
\end{figure}
\begin{figure}[t]
\begin{center}
\includegraphics[scale=0.46]{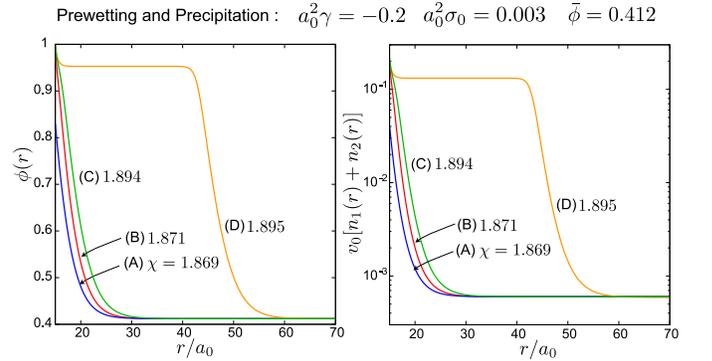}
\caption{(Color online) 
$\phi(r)$  (left) 
and   $v_0(n_1(r)+n_2(r))$ (right) 
at four points (A)-(D) marked in Fig.6 
for  a hydrophilic colloid  surface 
with $\gamma=-0.2a_0^{-2}$ and  $\bar{\phi}= 0.412$. 
 There is a weakly discontinuous 
 prewetting transition between 
(A) and (B) and a strongly discontinuous 
precipitation transition   between 
(C) and (D).
}
\end{center}
\end{figure}

\subsubsection{Hydrophobic surface with $\gamma=0.08a_0^{-2}$}

\begin{figure}[t]
\begin{center}
\includegraphics[scale=0.6]{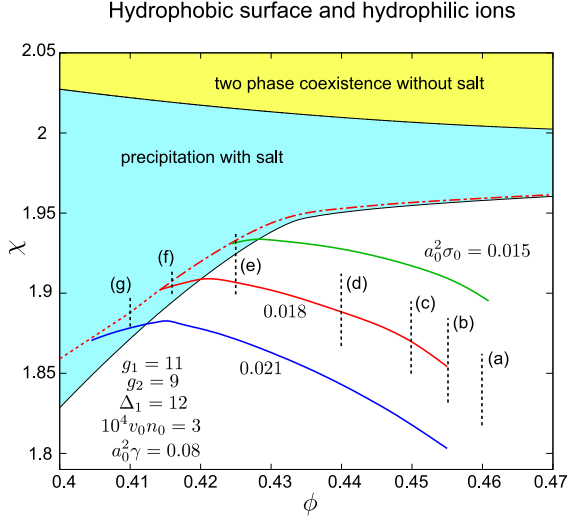}
\caption{(Color online) 
Phase diagram of  prewetting and precipitation 
transitions  on a hydrophobic colloid 
 with $a_0^2\gamma=0.08$. 
Three prewetting lines,  corresponding  to 
$a_0^2\sigma_0=0.015$, 0.018 and 0.021, 
are more extended than in the hydrophilic 
case in Fig.6. Precipitation on the colloid surface 
occurs above the broken red line. See  Fig.9  
for $\alpha$ and $\Gamma$ on vertical paths {(a)},$\dots$, 
and {(g)}.
}
\end{center}
\end{figure}

\begin{figure}[t]
\begin{center}
\includegraphics[scale=0.45]{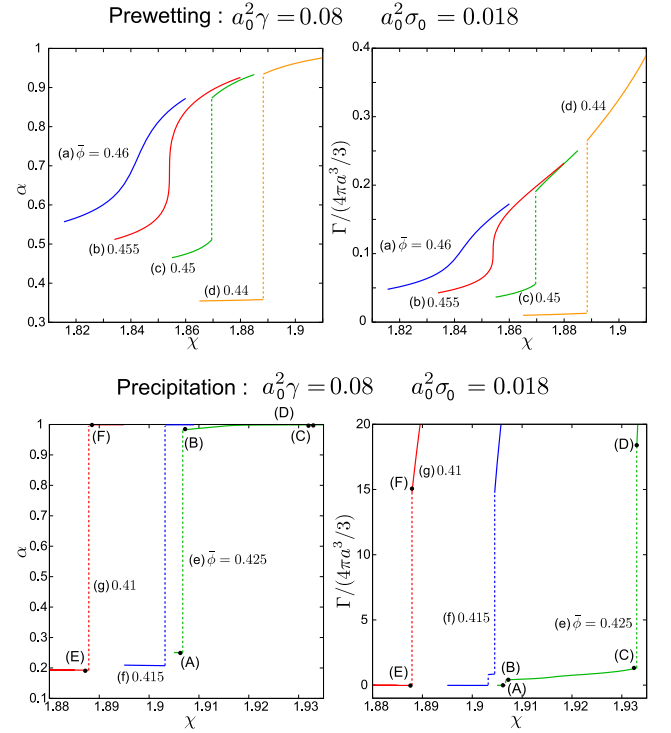}
\caption{(Color online) 
Degree of ionization $\alpha$ 
(left) and normalized  adsorption 
 $\Gamma/(4\pi a^3/3)$  (right) 
vs $\chi$ 
on a hydrophobic colloid with $a_0^2\gamma =0.08$ 
and $a_0^2\sigma_0=0.018$ along paths 
(a)   $\bar\phi=0.46$, (b) 0.455, (c) 0.45, (d) 0.44, (e) 0.425, 
(f) $0.415$, and (g) 0.41 (see their locations in Fig.8).   
Top: Behavior around the prewetting line 
at $a_0^2\sigma_0=0.018$ (middle line in Fig.8)  
along paths {(a)},{(b)}, {(c)}, and {(d)}. 
Bottom: Large jumps 
across the precipitation line  
along the paths {(e)},{(f)}, and {(g)} (dotted lines).
On path (e), there is a prewetting transition 
between (A) and (B) and a precipitation transition 
between (C) and (D). Path {(f)} also passes  the two 
transitions. Path {(g)}  passes only  the precipitation line. 
}
\end{center}
\end{figure}

\begin{figure}[t]
\begin{center}
\includegraphics[scale=0.45]{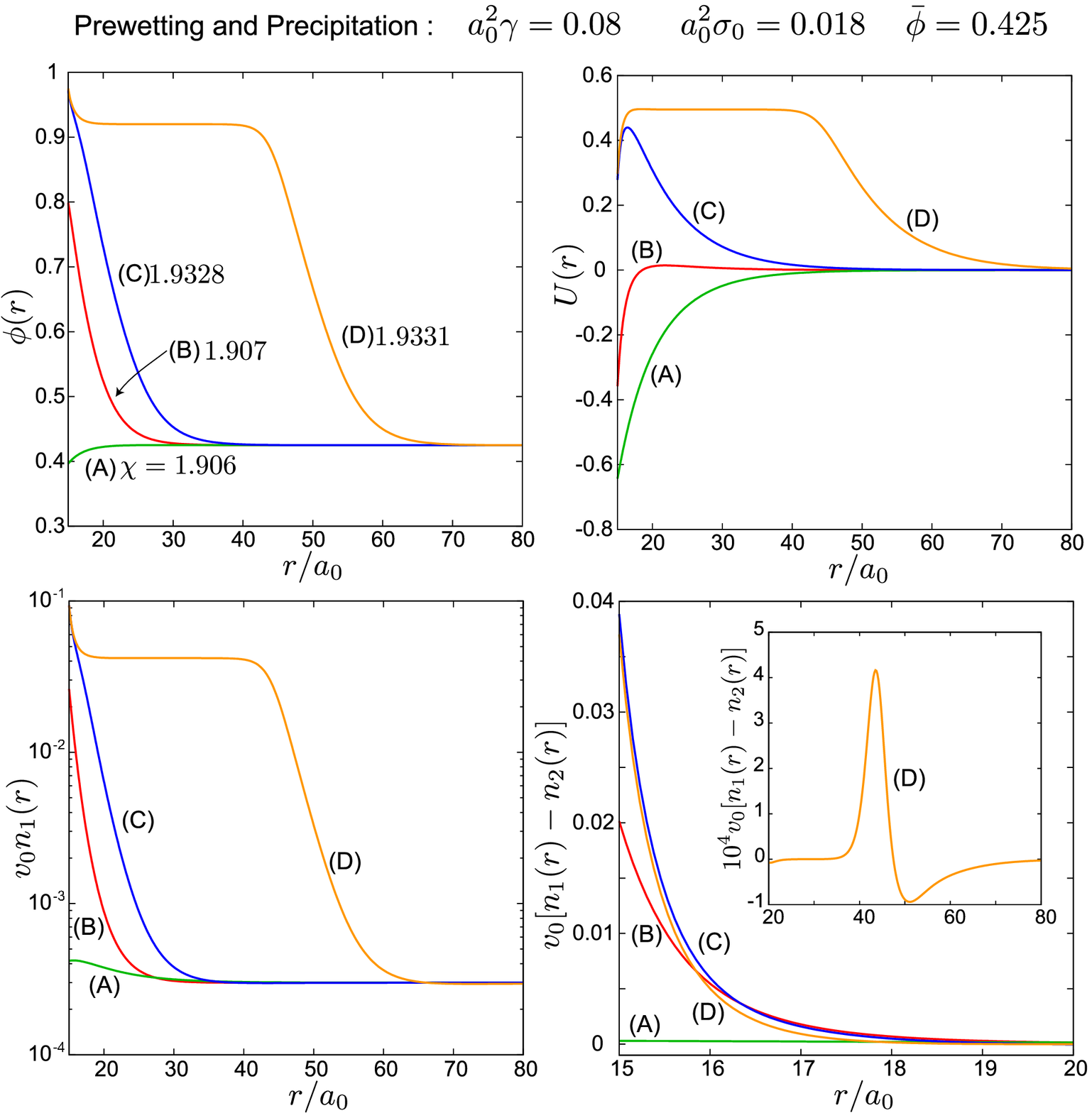}
\caption{(Color online) 
$\phi(r)$ (left top), $U(r)$ (right top), 
$v_0n_1(r)$ on a semi-logarithmic scale (left bottom), 
and $v_0(n_1(r)-n_2(r))$ (right bottom)  
on a hydrophobic colloid at four points (A),(B), (C), 
and (D) on path {(e)} in Fig.9, 
where  $a_0^2\gamma =0.08$,  
$a_0^2\sigma_0=0.018$, and $\bar\phi=0.425$. 
A prewetting transition occurs between 
(A) and (B), while  a precipitation transition 
between (C) and (D). At (B), (C), and (D), 
the ions are enriched in the layer. 
In the inset in the right bottom plate, 
 a small electric double layer is shown 
 in the interface region around $r\cong 40 a_0$ 
at (D).   
}
\end{center}
\end{figure}

The prewetting behavior is more exaggerated 
for a  weakly hydrophobic   surface  under Eq.(2.29) than  
for a  hydrophilic surface. 
In Subsec.IIC, we have discussed the changeover  from a 
hydrophobic to hydrophilic surface 
with progress of ionization.   
Here, 
 we  examine the prewetting and  precipitation 
 transitions   in the weakly hydrophobic 
case with $\gamma=0.08 a_0^{-2}$   in  Figs.8-17
and also in the strongly  hydrophobic 
case   with $\gamma=0.2 a_0^{-2}$   in  
the lower plates in Fig.12.

In  Fig.8, we display 
three examples  of the prewetting line   
corresponding  to $a^2\sigma_0=0.015$, 0.018, 
and 0.021, which are   pushed  downward with increasing 
$\sigma_0$. They even bend downward and much extend  
outside the bulk precipitation line. The precipitation line 
for a colloid (broken line) is inside 
the bulk precipitation region  and 
is  independent of $\sigma_0$ (as in  Fig.6). 
 
In Fig.9, we show the degree of ionization 
$\alpha$ and the excess adsorption  $\Gamma$ defined 
in Eq.(5.2) as functions of  $\chi$ for various $\bar\phi$. 
In the upper plates, the discontinuities across the prewetting 
line increase with increasing $\chi$, where   the 
 critical point  is located at the smallest $\chi$ 
 on the line.  (In sharp contrast, they 
 decrease with increasing $\chi$  in 
 the hydrophilic case  in Fig.6.) In the lower plates, 
  $\alpha$ and $\Gamma$ 
 are  shown   along  paths,  {(e), (f)}, and {(g)},  
 where {(e)} and {(f)}  pass  
 the prewetting and precipitation lines but path {(g)} 
 passes  the precipitation line only. 
In particular, on path {(e)}, $\alpha$ 
 changes from about 0.2 to values slightly smaller 
 than unity across the prewetting line or 
 between (A) and (B), while 
 $\Gamma$ changes greatly across the precipitation line 
 or  between (C) and (D).  
The  $\Gamma/(4\pi a^3/3)$ 
is $ -0.0137$ at (A), 
0.417 at (B), 1.30 at (C), and 19.1 at (D). 

The jump of $\Gamma$  between (C) and (D) 
is very large and can be explained by minimization of $F_{\rm wet}$ 
in Eq.(4.15) \cite{jump}.

In Fig.10,  the profiles of $\phi(r)$, $U(r)$, 
$n_1(r)$, and $n_1(r)-n_2(r)$  are shown 
at  points (A), (B), (C), and (D) on  path {(e)} 
in the right bottom panel of Fig.9. 
At  (A), which is  below the prewetting line, 
 the colloid surface is still hydrophobic 
 with  a negative surface value of $\phi'= 
 \p \phi/\p r$, 
 $U(r)$ is negative, and the ions are weakly  accumulated 
 near the surface. 
 At  (B), which is slightly above the prewetting line, 
 the surface is hydrophilic, 
 $U$ has a small maximum, 
 and the cations and the anions are both enriched near the 
 surface. 
 At (D) there is a thick wetting layer 
 enriched with ions.  In  the right bottom panel, 
 the charge accumulation 
  near the surface and the electric 
 double layer (in the inset) are shown, which are small 
 because of small difference $g_1-g_2=2$ in this case. 
 In passing, 
 the value of $\chi$ at the precipitation transition 
 is 1.9329,  
  while  it is  $1.9320$ for $d=25a_0$ (not shown here). 
These values  are  only 
slightly larger than $\chi_{\rm p}=1.9246$.

\subsection{Two colloids 
and their interaction free  energy 
for hydrophobic surface and hydrophilic ions}

Placing two hydrophobic  colloids along the $z$ axis, 
we now calculate  the    profiles of $\phi$ and $U$ 
and the interaction free energy $F_{\rm int}(d)$ as a function 
of the  distance $d$ between the two colloid centers in Figs.11-17. 
The profiles depend on  $r= ({x^2+y^2})^{1/2}$ and $z$. 
The degree of ionization 
$\alpha=\alpha(\theta)$ depends on the angle $\theta$ 
with respect to  the $z$ axis.  Its angle average 
is written as  
\be 
\av{\alpha}= 
\int_0^\pi  d \theta \sin\theta~ \alpha(\theta).
\en 
The precipitation transition is continuous 
for  the present parameters, 
as stated at the beginning of this section. 
  
\subsubsection{Crossover from 
repulsive to attractive interaction}

First, for $\bar\phi=0.425$,  we examine the 
behavior  around 
the prewetting  transition 
with  $a_0^2\gamma =0.08$ and  
$a_0^2\sigma_0=0.018$.    In Fig.11, the left panel  
gives  $F_{\rm int}$ vs $d$ 
for $\chi=1.88$ and 1.92. 
Remarkably, it is small and positive for $\chi=1.88$ 
but is  negative and is  much amplified  
 for $\chi=1.92$.  The  right  panel 
displays the corresponding $\phi(r,z)$   
at $d=44a_0=2.93a$, where 
$(\Gamma/(4\pi a^3/3), \av{\alpha})= 
(-0.0198,  0.305)$  for   $\chi=1.88$ 
and $(1.30, 0.941)$  for $\chi=1.92$ with a big difference 
in $\Gamma$.   The screening length 
 $\kappa^{-1}$  in Eq.(3.5)  is 
$7.94a_0 $,  while the correlation length $\bar\xi$ 
in Eq.(3.11) is 
$2.71a_0$ for $\chi=1.88$ 
and $3.23a_0$ for $\chi=1.92$.  
 The surface remains hydrophobic ($\phi'>0$)  
for $\chi=1.88$ but becomes  
 hydrophilic ($\phi'<0$) 
for $\chi=1.92$.

On the other hand,  at  the critical 
composition $\bar{\phi}=0.5$, 
there is   no prewetting transition, 
though the precipitation 
occurs for $\chi>\chi_{\rm p}= 1.9675$,  
as can be seen in Fig.4.  
The instability point $\tau=\tau_c$ 
occurs at  $\chi=1.97$ since $v_0 \tau_c= 0.06$ 
(see Eqs.(3.9) and (3.11)). 
Thus, at $\bar{\phi}=0.5$, 
the physical quantities change continuously 
with varying $\chi $ below $\chi_{\rm p}$.  
In Fig.12,   we  show  $F_{\rm int}$ vs $d$ 
in the left and  the corresponding $\phi(r,z)$   
at $d=44a_0=2.93a$ in the right for $\bar\phi=0.5$. 
The screening length 
 $\kappa^{-1}$ 
  is $8.14a_0 $, 
while the correlation length $\bar\xi$ 
 is $1.14a_0$ for $\chi=1.2$ 
and $14.1 a_0$ for $\chi=1.965$. 
(i) The upper panels are for 
the weakly hydrophobic case with 
$a_0^2\gamma =0.08$  and  
$a_0^2\sigma_0=0.018$, where   
$(\Gamma/(4\pi a^3/3), \av{\alpha})= 
(0.063,  0.65)$  for   $\chi=1.2$ 
and $(3.15, 0.98)$  for $\chi=1.965$.  
The surface is weakly hydrophilic at  $\chi=1.2$ 
and  is strongly  hydrophilic at  $\chi= 1.965$. 
(ii) The lower  plates are  
for the strongly hydrophobic case with 
$a_0^2\gamma =0.2$  and  
$a_0^2\sigma_0=0.01$, 
where 
$(\Gamma/(4\pi a^3/3), \av{\alpha})= 
(-0.093,  0.64)$  for   $\chi=1.2$ 
and $(-1.31, 0.32)$  for $\chi=1.965$.  
The adsorption of the oil component 
is enhanced on approaching the criticality. 

In the weakly hydrophobic case in the upper 
plates in Fig.12, 
the linear theory fairly holds 
at  $\chi=1.2$,  while  it breaks down at $\chi=1.965$.  
 In fact, $F_{\rm int}(d)/k_BT$ at $d\cong 2a$ 
 is  3.71 (linear theory) and 2.38 (nonlinear theory) 
for $\chi=1.2$, while it is -22.4 (linear theory) 
and -3.44 (nonlinear theory) for $\chi=1.965$. 
The surface value of  
   $\psi=\phi-{\bar\phi}$ is about 
0.02   for $\chi=1.2$ and 
 0.25 for $\chi=1.965$, while  $U\sim -1 $ both  
 for these cases.

\begin{figure}[t]
\begin{center} 
\includegraphics[scale=0.4]{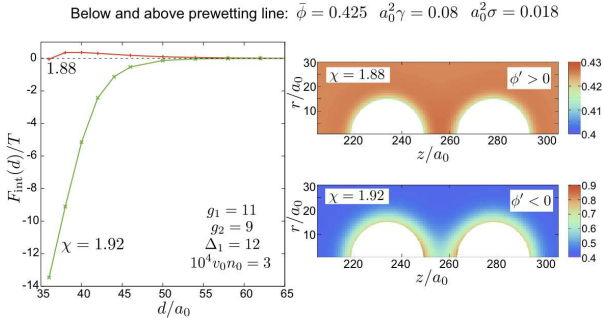}
\caption{(Color online)  
Normalized interaction 
free energy $F_{\rm int}(d)/T$ vs 
 normalized  separation distance 
$d/a_0$ (left) and $\phi(r,z)$ 
 at $d=44a_0=2.93a$ (right) for 
 $\bar\phi = 0.425$, where 
 $a_0^2\gamma =0.08$,  and 
 $a_0^2\sigma_0=0.018$.  
The system is below a prewetting line 
at  $\chi= 1.88$  and above it at  
$\chi= 1.92$, where   $F_{\rm int}$ 
changes its sign from positive to negative.  
The surface is hydrophobic with 
$\phi'= {\bi \nu}\cdot\nabla\phi>0$ (see Eq.(2.24)) 
 at $\chi=1.88$,  but   is hydrophilic  with 
$\phi'<0$ due to an increase in $\alpha$  at $\chi=1.92$.  
The color represents $\phi$ according to the color bar 
attached.  
}
\end{center}
\end{figure}

\begin{figure}[t]
\begin{center}
\includegraphics[scale=0.4]{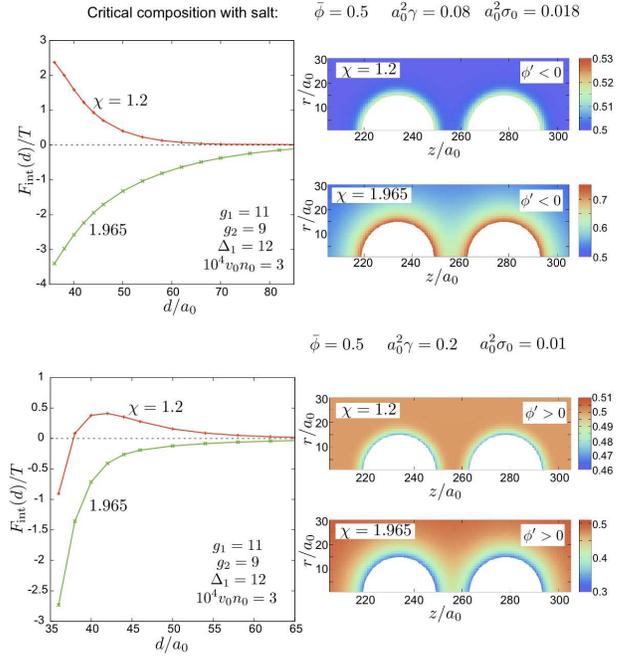}
\caption{(Color online) Normalized 
interaction free energy $F_{\rm int}(d)/k_BT$ vs 
 normalized  separation distance 
$d/a_0$ (left) and $\phi(r,z)$  at $d=44a_0=2.93a$   (right) 
at the critical composition $\bar\phi = 0.5$, 
where there is no prewetting transition. 
Top: Those  for (originally) weakly hydrophobic surface 
with $a_0^2\gamma=0.08$ and $a_0^2\sigma_0 =0.018$ 
at $\chi=1.2$  and $\chi=1.965$. 
Here $F_{\rm int}$ changes its sign 
on approaching the solvent criticality 
The surface becomes  hydrophilic ($\phi'<0$) 
weakly for $\chi=1.2$ and strongly  for $\chi=1.965$ 
 due to ionization. 
Bottom: Those  for  strongly 
hydrophobic surface  
with $a_0^2\gamma=0.2$ and $a_0^2\sigma_0 =0.01$ 
at $\chi=1.2$  and $\chi=1.965$. 
Here the surface remains hydrophobic  
even for full ionization.  
}
\end{center}
\end{figure}

\begin{figure}[t]
\begin{center}
\includegraphics[scale=0.4]{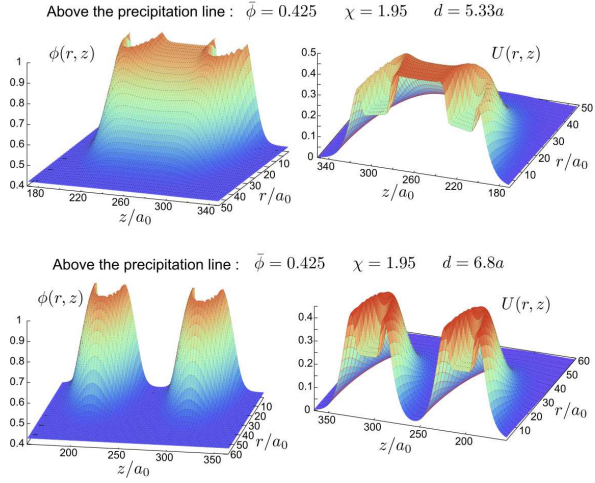}
\caption{(Color online) 
 $\phi(r,z)$ (left) and $U(r,z)$ (right) 
around two hydrophobic colloids 
above a  precipitation line,  
where $a_0^2\gamma=0.08$, 
 $a_0^2\sigma_0=0.018$, 
 and $\bar\phi=0.425$. 
  Wetting layers of two colloids are  
bridged at  $d=5.33a=80a_0$ (top) 
 and are disconnected for $6.8a=102a_0$ (bottom).  
The surface has become hydrophilic with 
$\phi'<0$.  
}
\end{center}
\end{figure}

\begin{figure}[t]
\begin{center}
\includegraphics[scale=0.58]{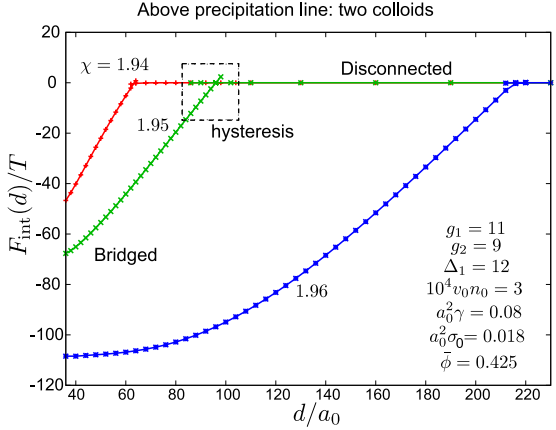}
\caption{(Color online) 
Normalized interaction free energy $F_{\rm int}(d)/T$ 
as a function of  normalized distance $d/a_0$ 
above a  precipitation line, where 
 $\chi=1.94$, $1.95$ and $1.96$ for three curves (from  above). 
Wetting layers of two colloids are bridged 
for   small  $d$ but are disconnected 
for large $d$. There is 
hysteretic behavior between these two states (see Fig.15). 
 }
\end{center}
\end{figure}

\begin{figure}[t]
\begin{center}
\includegraphics[scale=0.45]{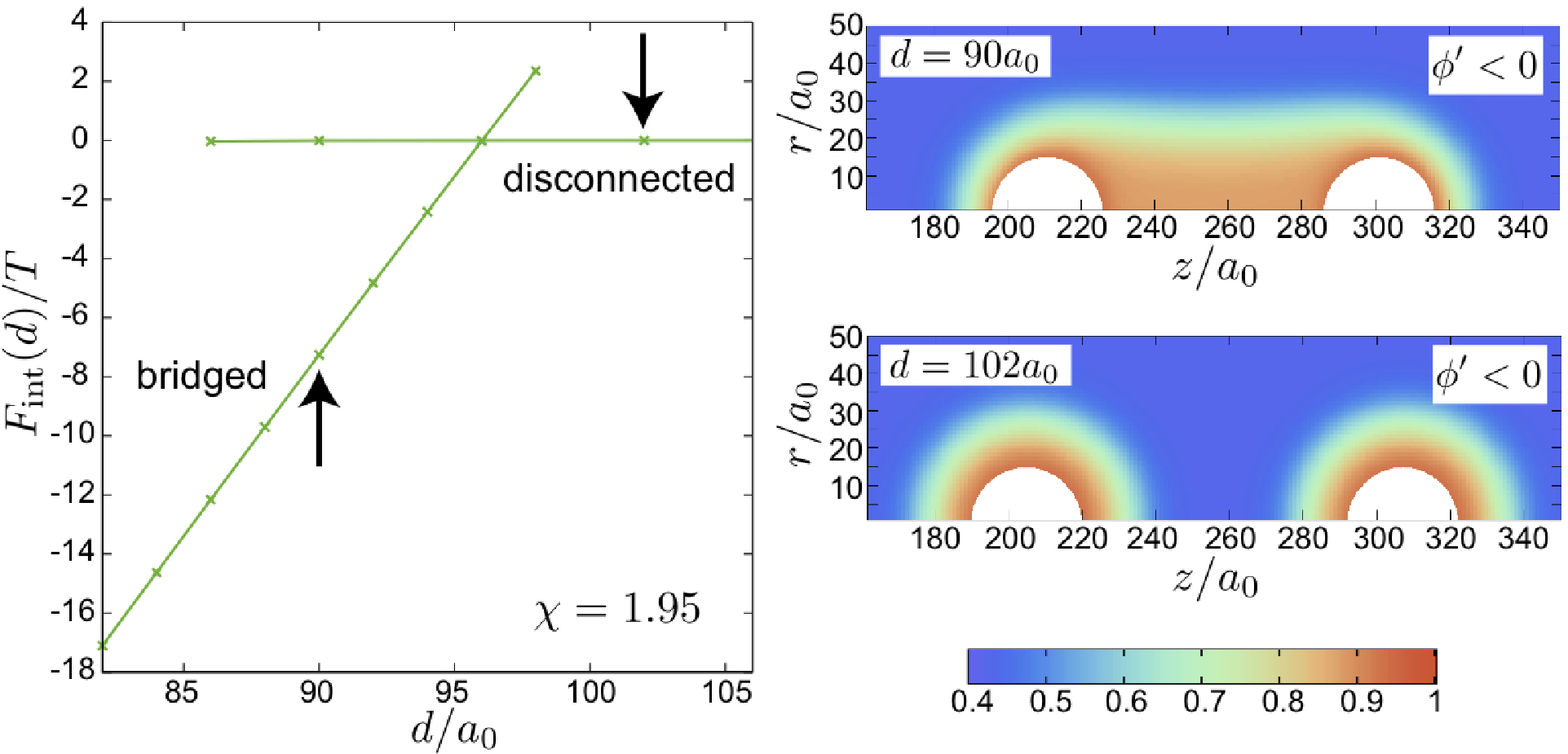}
\caption{(Color online) Left: Normalized 
interaction free energy $F_{\rm int}(d)/T$, exhibiting  
 hysteresis  between 
bridged and disconnected states of  wetting layers 
for $\chi=1.95$ in the box region in Fig.14. 
There can be a metastable 
state in an interval of $d$ (left).  Composition 
$\phi(r,z)$ in wetting layers 
 at the two states marked by the arrows in the left. 
}
\end{center}
\end{figure}

\subsubsection{ Bridged and disconnected  wetting layers}

Next, we examine the  wetting layer behavior 
 after the precipitation transition in Figs.13-15 for $\bar{\phi}=0.425$. 
In Fig.13, the profiles of $\phi(r,z)$  and $U(r,z)$ are  
presented  above the  precipitation  line  with 
the same parameter values as in Fig.11.
Here we have thick wetting layers 
on the two colloids and they are  
bridged for  $d=80a_0=5.33a$ (top plates) and 
are disconnected for $102a_0=6.8a$ (bottom plates).  
In Fig.14, 
the interaction free energy $F_{\rm int}(d)$ is shown 
as a function of  $d$, where 
the three curves correspond to 
 $\chi=1.94$, $1.95$,  and $1.96$. 
At  relatively short separation $d=44 a_0$,  for instance,  
$\Gamma$ increases 
with increasing $\chi$ as  
 $\Gamma/(4\pi a^3/3)=
3.87$, $7.20$, and $23.3$, respectively. 
We recognize that the wetting layers  are bridged 
for relatively  small  $d$ but are disconnected 
for large $d$,  exhibiting hysteresis. 
 The $F_{\rm int}(d)/T$  assumes  
negative values  of order $-100$ 
while bridged, 
but it becomes very small once detached. 
For larger colloid radii, 
 the value of $F_{\rm int}(d)/T$ 
should be much more amplified 
($\propto a^3)$. In Fig.15, the hysteretic 
transition behavior is illustrated   
between the bridged and disconnected sates. 
That is, in an interval of $d$, we find two linearly stable 
profiles, where  one is metastable 
with a higher   $F_{\rm int}(d)$. 
 
It is worth noting that  
Hopkins {\it et al.} \cite{Evans} also found a bridging 
of adsorption layers of two approaching neutral colloid 
particles  in  a mixture solvent 
close to the coexistence curve.

\subsubsection{Partial wetting on hydrophobic surface}

In the previous  examples 
of hydrophobic colloids,  
the surface  is in 
a completely wetted state above the precipitation line   
 in the composition range ${\bar \phi}=0.4-0.5$.  
However, for smaller ${\bar \phi}$, 
a  hydrophobic colloid 
surface can be 
 partially  
wetted if  $\gamma$  is not very large.

In Fig.16, at $\bar\phi=0.3$,   
we display such 
composition  profiles  $\phi(r,z)$  
around two hydrophobic colloids 
at $d=44a_0= 2.93a$ by changing $\chi$. 
The  other parameters are common 
to those in Figs.11 and 13.  For  $\chi>
\chi_{\rm p}=1.12$, there appears a water-rich 
region partially wetting  the colloid surface. 
The surface is   partially wetted for 
$\chi=1.25$, 1.55, and $1.85$, 
but is   completely   covered by the  
water-like component at $\chi= 2$. 
Here,  $\alpha$  in the non-wetted regions 
is  0.082, 0.058, and 0.051 
for $\chi=1.85$, 1.55, and 1.25, respectively, 
while $\alpha$ is nearly equal to unity 
in the wetted surface regions.  In these examples,  
the number of the counterions from 
the wetted surface 
is only about $0.4\%$ 
of the number of the cations 
in the wetting water-rich  region, 
where the latter is 
about $5600$ for $\chi=1.25$.

In Fig.17, we vary  $d$ at $\bar\phi=0.3$  
and $\chi=1.55$ 
to see how the  partial  
wetting of two hydrophobic colloids is changed 
between bridged and disconnected states. 
In the left  panel,  $F_{\rm int}(d)$ 
is displayed as a function of $d$, where   
hysteresis is exhibited  between these two states. 
In the right panel, 
we show $\phi(r,z)$ 
in the connected case at $d=60a_0$ 
and  in the disconnected case at at $d=72a_0$. 
In the latter case, a water-rich droplet 
is partially wetting the left colloid. 
In these cases, $\alpha$ 
is 0.99 in the wetted  part 
and about 0.05 in the nonwetted part.

\begin{figure}[t]
\begin{center}
\includegraphics[scale=0.38]{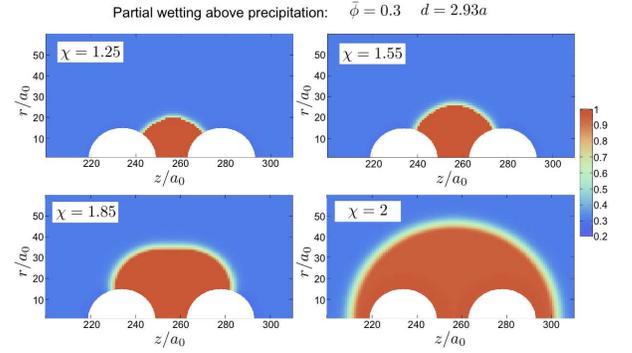}
\caption{(Color online) Partially or completely 
wetting profiles of $\phi(r,z)$  
around two hydrophobic colloids 
for $\bar\phi=0.3$  with  $d=44a_0= 2.93a$, where  
$\chi=1.25$, 1.55, 1.85, and 2 
above the bulk precipitation 
value $\chi_{\rm p}=1.12$, 
The  other parameters are common 
to those in Figs.11 and 13-15. 
}
\end{center}
\end{figure}

\begin{figure}[t]
\begin{center}
\includegraphics[scale=0.4]{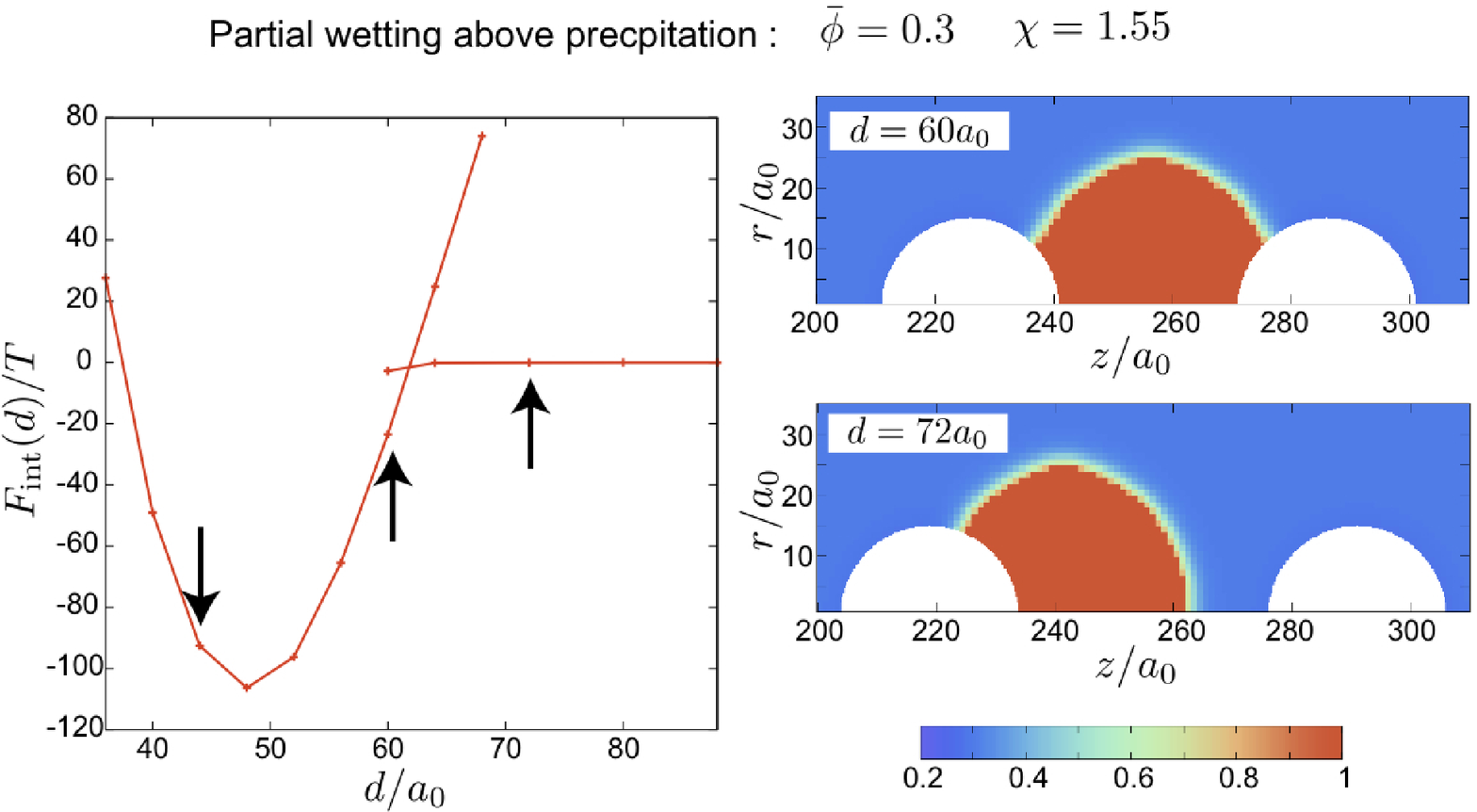}
\caption{(Color online) Two partially 
wetted, hydrophobic colloids in a  
bridged or  disconnected state for  $\bar\phi=0.3$ 
at $\chi=1.55$. The  other parameters are common 
to those in Fig.16.  
Left:  $F_{\rm int}(d)/k_BT$ vs $d/a_0$ with  
hysteresis  between the two states. 
Arrows indicate $d/a_0=44$, 60, 72. 
Right: $\phi(r,z)$  
at $d/a_0=60$ (bridged) and 72 (disconnected). 
Profile at  $d/a_0=44$ is in the top right panel in Fig.16. 
}
\end{center}
\end{figure}

\subsection{Antagonistic ions}

With  antagonistic ions added, 
oscillatory response can arise against 
local disturbances even if 
the system is in a homogeneous 
state in the bulk region, 
as  has been discussed in the linear theory 
in Sec.III.  Here, 
 we set $g_1=-g_2=13$,  $a_0^2\gamma=0.2$,  
 $a_0^2\sigma =0.016$, and 
 $\bar{\phi}=0.5$.

In Fig.18, we show  $\phi(r)$ and  $U(r)$ 
as functions of  $r/a_0$  
around a single colloid, 
where three curves correspond to 
 $\chi=1.961$, $1.965$,  and $1.966$. 
 Oscillatory behavior is 
 amplified with increasing $\chi$. 
 The system tends to a homogeneous  state 
 far from the colloid  for the smaller two $\chi$,  
 but it is in a periodically modulated phase 
 for the largest $\chi$, since  
     the homogeneous state is linearly unstable 
     for  $\chi>1.9651$. 
In Fig.19, we examine the case of two colloids, 
In its left panel, 
the interaction free energy $F_{\rm int}$ 
is plotted as a function of  $d$ for two colloids  
for $\chi=1.95$, $1.961$ and $1.965$. 
In its right panel, 
the profile of $\phi(r,z)$ is  given  
 for  $\chi=1.965$ 
at $d=49a_0=3.27a$, where 
a homogeneous state is linearly stable.

\begin{figure}[t]
\begin{center}
\includegraphics[scale=0.4]{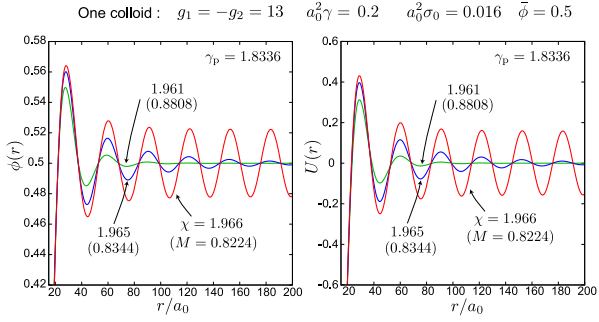}
\caption{(Color online) 
$\phi(r)$ and $U(r)$ around a  
hydrophobic colloid  
 for antagonistic salt with $g_1=-g_2=13$, 
where $a_0^2\gamma=0.2$,  
 $a_0^2\sigma_0  =0.016$, and $\bar{\phi}=0.5$. 
Three  curves correspond to 
 $\chi=1.961$ (green), $1.965$ (blue) and $1.966$ (red) 
 Parameter $M$ in Eq.(3.21) in the linear theory is 
 equal to 0.8808, 0.8344, 0.8224, respectively.  
 Oscillatory behavior is 
 amplified with increasing $\chi$. 
 The system tends to be homogeneous far from the colloid 
 for the first two curves,   while 
  a mesophase is realized for the third curve. 
 }
\end{center}
\end{figure}

\begin{figure}[t]
\begin{center}
\includegraphics[scale=0.43]{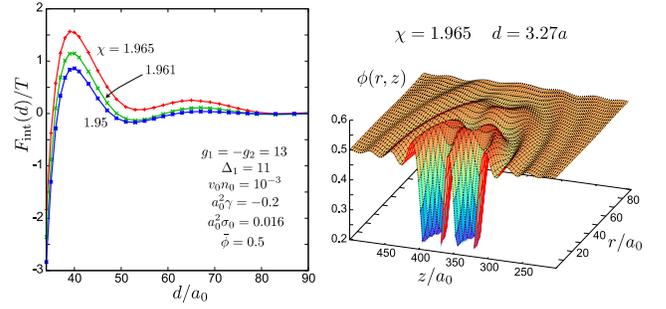}
\caption{(Color online) 
Left: Normalized interaction free energy $F_{\rm int}/k_BT$ 
as a function of  $d$ for two colloids  
with  antagonistic ions, where 
$a_0^2\gamma=0.2>0$, 
 $a_0^2\sigma_0=0.016$, 
 and $\bar\phi=0.5$. Three curves correspond 
 to $\chi=1.950$, $1.961$ and $1.965$ from below. 
Instability occurs at $\chi>1.9651$, 
so  homogeneity is attained 
far from the colloids. 
Right: $\phi(r,z)$ around two hydrophobic colloids 
 at $d=49a_0=3.27a$   for  $\chi=1.965$.     
 }
\end{center}
\end{figure}

\section{Summary and remarks}
\setcounter{equation}{0}

In summary, we  have 
examined how ionizable  colloids 
influence  the ion distributions and the composition 
field in binary polar solvents. These perturbations then 
give rise to the interaction free energy $F_{\rm int}(d)$ 
between  two colloids as a function of  the 
distance $d$ between their  centers.
We summarize our main results.\\ 
(i) In Sec.II, we have introduced 
a  Ginzburg-Landau model  in the presence of 
negatively ionizable colloids. The 
 fundamental fluctuating variables are 
 the composition $\phi$,  the ion 
densities $n_i$, and the degree of ionization $\alpha$, 
which are inseparably coupled in the presence 
of the selective  solvation. 
Important parameters in the bulk free energy 
are the interaction parameter $\chi$ 
(determined by the temperature $T$), 
the average composition $\bar\phi$, 
the average anion density $n_0$, and 
the solvation parameters $g_i$. 
Those related to the colloid  are 
the radius $a$, 
the molecular interaction parameter 
$\gamma$ representing the surface field, 
the density of the ionizable groups on the surface $\sigma_0$, 
and the composition-dependent  
ionization free energy $T(\Delta_0-\Delta_1\phi)$, which 
 appear in the boundary condition (2.24) 
on $\phi$ and the mass action law (2.26) for $\alpha$. 
\\ 
(ii) In Sec.III, we have presented 
a linear theory of the electrostatic 
and compositional disturbances 
produced by charged colloids, which is a generalization of 
the Debye-H$\ddot{\rm u}$ckel 
and  DLVO theories. In the linear scheme, the colloid interaction 
free energy $F_{\rm int}(d)$ is a linear combinations 
of two Yukawa functions 
$e^{-q_i (d-2a)}/(1+q_i/a)^2d$ as a function of 
the colloid separation distance $d$ 
in terms of two characteristic wave numbers $q_i$. 
In the weak coupling linit $g_a=g_1-g_2 
\to 0$, they tend to the DLVO interaction 
in Eq.(1.2) and the adsorption-induced  attraction 
in Eq.(1.5).\\
(iii) In Sec.IV, we have presented a theory of 
precipitation  on the colloid surface assuming a 
completely  wetting layer, 
which is induced by the selective solvation 
of hydrophilic ions far from the solvent coexistence curve. 
This precipitation occurs at small $\chi$ for 
relatively small $\bar\phi$ (say, 0.1) 
(see the left panel of Fig.4). 
However, the growth of the layer thickness 
is slow with increasing $\chi$ for small $\bar\phi$. 
For ${\bar\phi}\gs \phi_c$ 
precipitation occurs close to the solvent coexistence curve. 
\\
(iv) In Sec.V, we have presented 
numerical results on precipitation and prewetting 
on the colloid surface 
for  hydrophilic $(\gamma<0$) 
and hydrophobic $(\gamma>0$)  surfaces  in the nonlinear theory. 
We are particularly interested in 
the weakly hydrophobic 
surface satisfying Eq.(2.29). Such a surface 
is hydrophobic without ionization, but becomes hydrophilic 
with progress of ionization. Also 
the prewetting phase transition is 
more dramatic for such a hydrophobic surface 
than for a hydrophilic surface as in Figs.6 and 8. 
Wetting layer formation occurs above 
a precipitation line, which   
weakly depends  on the radius $a$ and is located 
slightly above the bulk precipitation 
line $\chi=\chi_{\rm p}$.
Such layers  undergo a bridging 
transition with a great change in 
the  interaction free energy 
$F_{\rm int}$ as in Figs.13-15. 
They  either completely or partially wet 
the surface 
depending on the average 
composition $\bar \phi$ as in Figs.16-18.  
For antagonistic ion pairs, oscillation can be seen  
the composition and potential 
profiles as a   function of the separation 
distance $d$ as in Fig.18, 
but it is largely masked in  
$F_{\rm int}$ in Fig.19.
\\

We make some  remarks.\\
1) Our coarse-grained theory is inaccurate 
on the angstrom scale, but the solvation parameters 
$g_i$ and the ionization parameter $\Delta_1$ 
can be made very large, so  the precipitation and 
prewetting transitions on the colloid 
surface  have been  predicted. 
We also note that   the molecular volumes 
of the two components are often very different in real 
mixtures.  For example, those  of 
 D$_{2}$O and tri-methylpyridine   (the inverse densities of 
the pure components) are 28 and 
 168 \AA$^3$, respectively \cite{Sadakane}. 
The coefficient $C$ in Eq.(2.1) 
of the gradient free energy 
remains  an arbitrary constant \cite{comment-gra}, 
though we have set  $C= 2a_0^{-1}$ in Sec.V. 
\\ 
2)  We believe  that 
the previous observations of 
 colloid  aggregation  \cite{Beysens,Maher,Kaler,Guo,Bonn}   
 should be induced by overlapping 
of enhanced adsorption or wetting layers 
on the colloid surface \cite{Evans}. 
If  the ions are neglected, 
  colloid particles constitue a 
selective   solute added  in 
 a binary mixture \cite{Kaler}. Furthermore, 
 if the selectivity is high, 
 our previous theory \cite{Okamoto} 
indicates a solute-induced phase separation 
 with a phase  diagram as in Fig.4. 
This aspect should be studied in more detail.  
\\
3) 
Wetting behavior remains unexplored 
in the presence of a highly selective 
solute.  It becomes even more complex 
if the substrate itself is ionizable. 
We have realized both complete and partial wetting 
on ionizable colloid surfaces, but 
the information gained is still fragmentary 
because many parameters are involved in 
the problem. 
\\
4) The effects of the critical fluctuations 
on the interactions between  solid surfaces 
are very intriguing \cite{Fisher,Krech,JSP,Nature2008}. 
 Ions should further  promote bridging 
 of highly adsorbing or  wetting layers.\\
5) 
For antagonistic ion pairs, 
the oscillatory behavior in the  
colloid interaction in Fig.19 is rather mild, though 
it is evident in the composition and potential profiles. 
It becomes  more evident in the interaction 
between two parallel plates, as in the case of 
liquid crystals \cite{Uchida}.\\ 
6) 
For polyelectrolytes including ionized gels, 
there are a number of unsolved problems arising from 
selective solvation. Even in one-component solvents, 
ions interact differently with polymer segments  and solvent 
molecules \cite{Onuki-Okamoto}. 
In mixture solvents (water-alcohol), 
a wetting film  should be formed around a chain, 
as stated in Sec.I  
\cite{Onuki-Okamoto,B1,B2,B3}. 
The  Manning-Oosawa counterion condensation 
mechanism should be modified 
for mixture solvents.

\begin{acknowledgments}
This work was supported by Grant-in-Aid 
for Scientific Research on Priority 
Area ``Soft Matter Physics'' from the Ministry of Educ
ation, Culture, Sports, Science and Technology of Japan. 
One of the authors (A.O.) would like to thank 
D. Bonn, D. Beysens,  and   H. Ohshima, 
for informative correspondence. 
\end{acknowledgments}

\vspace{2mm} 
\noindent{\bf Appendix A: 
Pair correlation functions}\\
\setcounter{equation}{0}
\renewcommand{\theequation}{A\arabic{equation}}


\subsubsection{Composition fluctuations}

We examine the structure factor  of the composition 
fluctuations  
$S(q)
= {\av{|\phi_{\small{\bi q}}|^2}}_{\rm e}$, where
$\phi_{\small{\bi q}}$ is the Fourier component 
of $\phi({\bi r})$ with wave vector $\bi q$ 
and  $\av{\cdots}_{\rm e}$ 
denotes taking the thermal average. 
The mean-field structure factor   reads 
\cite{Onuki-Kitamura,OnukiPRE},    
\be
\frac{1}{ S(q)}= \tau-\tau_c  + C{q^2}\bigg [1-  
\frac{\gamma_{\rm p}^2 \kappa^2}{q^2+\kappa^2}\bigg] , 
\en 
in terms of $\tau$ in Eq.(3.6), $\tau_c$ in Eq.(3.9),  
and $\gamma_{\rm p}$  in Eq.(3.22).  
If the right hand side of Eq.(A1) is expanded  
with respect to $q^2$, the coefficient in front of 
$q^2$ is $C(1-\gamma_{\rm p}^2)$. 
Thus  a Lifshitz point is realized at $\gamma_{\rm p}=1$. 
(i) For $\gamma_{\rm p}<1$, $S(q)$ is maximum 
at $q=0$ and $S(q)^{-1} \propto \bar{\xi}^{-2} 
+(1-\gamma_{\rm p}^2)q^2$ 
for  $q\ll \kappa$, so 
a thermodynamic    instability occurs  for $\tau < \tau_c$ 
at long wavelengths. To be precise, $\tau_c$ is the shift 
in this case. 
(ii) For $\gamma_{\rm p}>1$,  $S(q)$ has a peak at 
\be 
 q_{\rm  m}=( \gamma_{\rm p}-1)^{1/2}\kappa.
\en 
The corresponding   peak height is given by 
\be 
 S(q_{\rm m})=C^{-1}\kappa^{-2}/[M^2  
 - ( \gamma_{\rm p}-1)^{2} ]. 
\en  
which diverges as $ \gamma_{\rm p}-1 \to 
M= (\kappa\bar{\xi})^{-1}$. 
A mesophase (a charge-density-wave 
phase) should emerge   for $ \gamma_{\rm p}-1>M$, 
as was observed experimentally \cite{Sadakane}.

Furthermore, from Eq.(A1),  the quadratic equation 
(3.19) is identical to  $S(q)^{-1}=0$ 
with  $q^2=-\kappa^2\lambda$. 
In terms of $q_1$ and $q_2$ in Eq.(3.18), we obtain  
\be 
CS(q)= \frac{q^2+\kappa^2}{(q^2+q_1^2)(q^2+q_2^2)}.
\en 
The inverse  Fourier transformation of $S(q)$
yields the pair correlation $g(r)= 
\av{\delta\phi({\bi r})\delta\phi({\bi 0})}_{\rm e}$ 
for the composition fluctuations. It follows 
 a sum of the two Yukawa functions, 
\be 
g(r)=  \frac{(\lambda_1-1) e^{-q_1r}-(\lambda_2-1) 
e^{-q_2r}}{4\pi C(\lambda_1-\lambda_2)r}.
\en 
In particular, in the region $|M-1|<\gamma_{\rm p}<M+1$,  
$q_1$ and $q_2$ are complex conjugates and $g(r)$ behaves as 
\be 
g(r)= \frac{e^{-q_R r}}{4\pi Cr}\bigg 
[\cos (q_I r) +({q_a^2-q_I^2-\kappa^2})
 \frac{\sin (q_Ir)}{2q_R q_I} \bigg ], 
\en 
where $q_R$ and $q_I$ are are given in Eq.(3.24). 

\subsubsection{Ion fluctuations}
We eliminate  
 the composition  fluctuations 
 assuming their Gaussian distribution, where 
   the ion densities are  held fixed. 
The resultant   ion-ion 
potentials read \cite{Onuki-Kitamura,OnukiPRE}  
\be 
{V_{ij}(r)}= Z_iZ_j 
\frac{e^2}{\ve r}-\frac{T g_ig_j }{4\pi C}
\frac{1}{r}{e^{-r/\xi}} ,
\en 
where $Z_1$ and $Z_2$ are  $\pm 1$ in the monovalent case. 
The second term  is the composition-induced interaction 
   decaying  exponentially with    
 the correlation length   $\xi= (C/{\tau})^{1/2}$.   
It is attractive among the ions of the same species $(i=j)$. 
 It  dominates   over the  Coulomb 
repulsion  for $g_i^2> 4\pi C \ell_B$ 
in the range $r\ls \xi$, under which 
there should be a  tendency of ion  aggregation. 
 In the antagonistic case 
($g_1g_2<0$), the cations and anions 
 repel one another  for
$|g_1g_2|> 4\pi C \ell_B$  in the range $r\ls \xi$,   
leading to charge-density-wave formation  near 
the  criticality. 
Note that the shifted correlation length $\bar\xi$ 
 in Eq.(3.11) has appeared in the colloid-colloid interaction,   
 where both the composition and ion densities 
 are eliminated.

In our recent review papers \cite{Current,Nara}, 
we have furthermore calculated 
the structure factors  among the ion densities 
$G_{ij}(q)= 
{\av{n_{i\small{\bi q}}n_{j\small{\bi q}}^*}}_{\rm e}/n_0$. 
Further using Eq.(A4) we find 
\bea 
&&\hspace{-3mm}{G_{11}(q)}=1-\frac{1/2}{u+1} 
+ \frac{\gamma_{\rm p}^2[u+w(u+1)]^2/2}{(u+1)(u+\lambda_1)(u+\lambda_2)} ,
\nonumber\\
&&\hspace{-3mm}{G_{12}(q)}=\frac{1/2}{u+1} +  
 \frac{\gamma_{\rm p}^2[w^2(u+1)^2-u^2]/2}{(u+1)(u+\lambda_1)(u+\lambda_2)},
\ena 
where $u= q^2/\kappa^2$ and $w=(g_1+g_2)/(g_1-g_2)$. 
The $G_{22}(q)$ is obtained   if $w$ in $G_{11}(q)$ 
is replaced $-w$.
The inverse Fourier transformation 
of these structure factors 
gives rise to the pair   correlation functions  
$g_{ij}(r)=\av{\delta n_i({\bi r}) 
\delta n_j({\bi 0})}_{\rm e}/n_0$. 
We notice that the terms  proportional to $e^{-\kappa r}/r$ 
cancel to vanish from  the relation $(1-\lambda_1)
(1-\lambda_2)=\gamma_{\rm p}^2$. Thus,    
\be 
g_{ij}(r)
= {\delta ({\bi r})}\delta_{ij} - 
(K_1^{ij} {e^{-q_1r}} + K_2^{ij}{ e^{-q_2r}}) \frac{1}{r} ,
\en 
where   the $\delta$ function 
appears due to the self correlation 
and $K_1^{ij}$ and $ K_2^{ij}$ are 
appropriately defined constants.

\vspace{2mm} 
\noindent{\bf Appendix B: Calculations 
in the linear theory for one and two colloid 
particles}\\
\setcounter{equation}{0}
\renewcommand{\theequation}{B\arabic{equation}}

First, we seek  
the fundamental profiles $U_0(r)$ and $\psi_0(r)$ 
around a single colloid particle induced by 
by the boundary conditions (3.12) and 83.13), 
where the surface charge is fixed. 
They depend only on the distance $r$ from the colloid center. 
For $r>a$ we  may set 
\bea 
&&U_0(r)= g_a (\alpha_1 {e^{-q_1 r}}- \alpha_2 
{e^{-q_2r}})/{r},\\ 
&&\psi_0 (r) =(\beta_1 e^{-q_1 r}- 
\beta_2 {e^{-q_2r}})/{r}.
\ena 
For $r<a$,  
we have $U_0(r)= U_0(a)=$const. from $\nabla^2U_0=0$. 
 From   Eqs.(3.7), (3.8), and  (3.21), we obtain   
\bea 
&&  \beta_i =(1-\lambda_i) \alpha_i, \\
&&[\lambda_i-M^{2}+ \gamma_{\rm p}^2]  
\beta_i=  \gamma_{\rm p}^2\alpha_i,
\ena  
which hold  for $i=1,2$. 
The boundary conditions (3.12) and (3.13) give 
\bea 
&&\alpha_1' - \alpha_2'= -4\pi \ell_B {\bar \sigma}/g_a ,\\
&&\alpha_1'(1-\lambda_1) - 
\alpha_2'(1-\lambda_2) = -{\bar\gamma}/C ,
\ena
where $\alpha_i'= \alpha_i (1+q_i a)e^{-q_i a}/a^2$. 
Using the relation $(\lambda_1-1)(\lambda_2-1)=
\gamma_{\rm p}^2$, we solve these 
equations to obtain   
\be 
\alpha_i= \frac{e^{q_i a} B_i(\lambda_1-\lambda_2)}{(
{1+q_i a })(1-\lambda_i)}, 
\en 
where $B_1$ and $B_2$ are defined in Eq.(3.26). 
We thus  confirm   Eqs.(3.25) and (3.26). 
In this one-colloid case $\Delta \Omega$ in Eq.(3.17) 
is written as ${F_{\rm self}}$. Some calculations give      
\be 
{F_{\rm self}}= {2\pi a^3T} 
\bigg[\frac{E_1}{1+ q_1 a}
- \frac{E_2}{1+ q_2a}\bigg] ,
\en 
where  $E_1$ and $E_2$  are defined in  Eq.(3.31).

Next we consider two colloid particles at positions 
${\bi r}_1$ and ${\bi r}_2$  separated by  
$d=|{\bi r}_1- {\bi r}_2|>2a$ under the condition of 
 fixed surface charge. 
In the colloid exterior ($|{\bi r}- {\bi r}_1|>a$ 
and $|{\bi r}- {\bi r}_2|>a$),  $U$ and $\psi$ are expressed as 
\bea
&&U = U_0(|{\bi r}-{\bi r}_1|) +
 U_0(|{\bi r}-{\bi r}_2|) + W ({\bi r}), \\
&&\psi = \psi_0(|{\bi r}-{\bi r}_1|) +
 \psi_0(|{\bi r}-{\bi r}_2|) + \Psi ({\bi r}),  
\ena 
where $U_0$ and $\psi_0$ are the fundamental profiles 
for   a single colloid. 
In the colloid exterior, we expand 
the corrections $W$ and $\Psi$ 
around the center  of the first colloid  at ${\bi r}_1$ as 
\bea 
&&W= g_a \sum_{i,\ell} a_{i\ell } 
k_\ell(q_i|{\bi r}-{\bi r}_1|) 
P_{\ell}(\theta_1), \\
&&\Psi= \sum_{i,\ell} (1-\lambda_i) 
a_{i\ell } k_\ell(q_i|{\bi r}-{\bi r}_1|) 
P_{\ell}(\theta_1), 
\ena 
where $ a_{i\ell }$ ($i=1,2$ and $\ell=0, 1, \cdots)$  
are unknown coefficients to be determined below. 
 The $P_\ell(\theta_1)$ 
are  the spherical harmonic functions 
with $\theta_1$ being the angle between 
${\bi r}-{\bi r}_1$ and ${\bi r}_2-{\bi r}_1$. 
We introduce the modified spherical 
Bessel functions  $i_\ell(x)$ and $k_\ell(x)$ \cite{Watson}.  
They  satisfy  
$i_\ell''+ 2i_\ell'/x -\ell(\ell+1)i_\ell/x^2 
=i_\ell$ and  $k_\ell''+ 2k_\ell'/x -\ell(\ell+1)
k_\ell/x^2 =k_\ell$, where we write $i_\ell'= di_\ell/dx$,  
 $k_\ell'= dk_\ell/dx$,  $i_\ell''= d^2i_\ell/dx^2$, 
  and $k_\ell''= d^2k_\ell/dx^2$. 
We have   
$i_\ell \sim x^\ell$ as $x\to 0$ 
and  $k_\ell \sim e^{-x}$ as $x\to \infty$.  
In particular \cite{Bessel}, 
\be 
i_0(x)= \sinh x/x, \quad k_0(x)= e^{-x}/x.
\en  
Thus, $i_\ell P_\ell$ and $k_\ell P_\ell$ 
satisfy the Helmholtz equations, 
\bea 
&&\hspace{-5mm}
(\nabla^2-q_i^2)[i_\ell (q_i|{\bi r}-{\bi r}_1|)
P_\ell (\theta_1) ]=0,\nonumber\\
&&\hspace{-5mm}
(\nabla^2-q_i^2)[k_\ell (q_i|{\bi r}-{\bi r}_1|)
P_\ell (\theta_1) ]=0.
\ena 
With   these relations and Eq.(3.7), 
we derive Eq.(B12)  from Eq.(B11).
In Eqs.(B9) and (B10) we also  need to 
expand $U_0(|{\bi r}-{\bi r}_2|)$ 
and $\psi_0(|{\bi r}-{\bi r}_2|)$ around ${\bi r}_1$ 
in terms of $P_\ell(\theta_1)$.   
To this end, we  use   the following 
mathematical relation \cite{Ohshima,Watson}, 
\be 
\hspace{-0.5mm}\frac{e^{- q|{\bi r}-{\bi r}_2|}}{|{\bi r}-{\bi r}_2|}
=\sum_\ell (2\ell+1) q k_\ell (qd)
i_\ell (q|{\bi r}-{\bi r}_1|)P_{\ell}(\theta_1),  
\en 
which holds for Re$(q)>0$ and
in the region $|{\bi r}-{\bi r}_1|<d=|{\bi r}_1- {\bi r}_2|$.   
On the other hand, 
in the  interior of the first colloid 
$|{\bi r}- {\bi r}_1|<a$, 
we have $\nabla^2U=0$ so that we may assume the expansion,  
\be 
U= U_0(a)+ g_a \sum_\ell b_\ell 
|{\bi r}-{\bi r}_1|^\ell P_{\ell}(\theta_1).   
\en

We can calculate the coefficients $a_{i\ell}$ and 
$b_\ell$ from the boundary conditions (3.12) and (3.13) 
and the continuity of $U$ at the colloid surface. 
We are interested in 
 the free energy deviation $\Delta \Omega$ in 
Eq.(3.17). For two symmetric colloids, we obtain  
\be 
{\Delta \Omega}/T= 
4\pi a^2 [ {\bar\sigma}\av{\psi}_0- \bar{\gamma}\av{U}_0],
\en  
where $\av{\cdots}_0$ denotes taking 
the surface average on the  colloid 1  
($= \int_0^\pi 
 d\theta_1 \sin\theta_1 (\cdots)/2$ at 
$|{\bi r}- {\bi r}_1|= a$).  
Thus $\Delta \Omega$ 
arises from the terms with $\ell=0$ in Eqs.(B11), 
(B12), and (B15).  For $\ell=0$, 
the boundary conditions 
(3.12) and (3.13) simply yield  
\bea 
&&\hspace{-8mm} 
\sum_i q_i [ \alpha_{i0} q_i k_0(q_id)i_{i0}' 
+  a_{i0} k_{i0}' ]=0, \\
&&\hspace{-8mm} 
\sum_i (1-\lambda_i)q_i
[ \alpha_{i0}q_i k_0(q_id)i_{i0}' 
+  a_{i0} k_{i0}' ]=0.
\ena 
For simplicity, we write 
$i_0(q_i a)$, 
 $i_0'(q_i a)$, $k_0(q_i a)$, 
 and  $k_0'(q_i a)$  as 
$i_{i0}$,  $i_{i0}'$, $k_{i0}$, 
 and  $k_{i0}'$, respectively, suppressing $q_ia$. 
In Eq.(B18) there is no contribution  
from the electric field within the colloid ($\propto \ve_{\rm p}$).
This is because the angle average 
of $({\bi r}-{\bi r}_1)\cdot \bi{E}$ vanishes 
 from Eq.(B16).     For each $i$, it follows 
 the relation, 
\be 
a_{i0}= -\alpha_i q_i k_0(q_i d) i_{i0}'/k_{i0}'.   
\en 
Elimination of $a_{i0}$ yields 
\be
{\av{U}_0}
=U_0(a)+ g_a\sum_i\alpha_i k_0(q_i d)
[i_{i0} -k_{i0}i_{i0}'/k_{i0}']. 
\en 
The expression for ${\av{\psi}_0}$ also follows  
in the same manner.  
Further, using  the relation, 
\be 
i_0 (x)- i_0'( x)k_0(x)/k_0'(x)= e^x/(1+x),
\en 
we  obtain   simple Yukawa forms,  
\bea
&&\hspace{-3mm}{\av{U}_0}
=U_0(a)+ g_a\sum_i\alpha_i 
\frac{e^{-q_i (d-a)}}{(1+q_ia )d}, 
\\
&&\hspace{-3mm}{\av{\psi}}_0
=\psi_0(a)+ \sum_i\alpha_i 
\frac{(1-\lambda_i) e^{-q_i (d-a)}}{(1+q_ia)d}  .
\ena  
We also have $b_0= [\av{U}_0-U_0(a)]/g_a$ from the 
continuity of $U$.  
Substitution of Eqs.(B23) and (B24) into 
Eq.(B17) leads  to the interaction free energy 
\be 
 F_{\rm int}= \Delta \Omega- 2 F_{\rm self}, 
\en 
given in Eq.(3.30).

Finally, we calculate the terms with  $\ell \ge 1$, 
though they do not contribute to $\Delta \Omega$ 
in the linear theory. 
From Eqs.(3.12) and (3.13) 
we express $a_{i\ell}$ in terms of  $b_\ell$ and $\alpha_i$ as 
\be
a_{i\ell}=[\ell b_\ell a^{\ell-1}\mu_i\ve_{\rm p}/\bar{\ve}
 -\alpha_i q_i^2 k_\ell(q_i d) i_{i \ell }']/q_i k_{i\ell }',
\en 
where $\mu_1= (1-\lambda_2)/(\lambda_1-\lambda_2)$
and  $\mu_2= (1-\lambda_1)/(\lambda_2-\lambda_1)$ in 
the first term.
We write  $i_{i\ell}= i_\ell(q_i a)$, 
 $i'_{i\ell}= i_\ell'(q_i a)$, 
  $k_{i\ell}= k_\ell(q_i a)$, 
 and $k'_{i\ell}= k_\ell'(q_i a)$.  
Requiring  the continuity of the potential, we determine  $b_\ell$ 
in the form,  
\be
b_\ell a^{\ell-1}= \frac{\sum_i 
\alpha_i q_i k_\ell(q_i d) \eta_{i\ell}}{a 
- \ell(\ve_{\rm p}/\bar{\ve}) \sum_i \mu_i w_{i\ell} /q_i },
\en 
where $\eta_{i\ell}= i_{i\ell}- i_{i\ell}'k_{i\ell}/k_{i\ell}'$ 
 and $w_{i\ell}=
  k_{i\ell}/k_{i\ell}'$. The term 
  proportional to $\ve_{\rm p}/\bar{\ve}$ in the denominator 
 in the right hand side arises from the boundary condition (3.12).


\begin{thebibliography}{99}


\bibitem{Dej} B.V. Derjaguin 
 and L.D. Landau, 
 Acta Physicochim.(USSR), {\bf 14}, 633 (1941). 


\bibitem{Ov} E.J.W.  Verwey  
and J.Th.G. Overbeek, 
 {\it Theory of the Stability of Lyophobic Colloids} 
 (Elsevier, Amsterdam, 1948). 

\bibitem{Russel} 
W. B. Russel, D. A. Saville, 
and W.  R. Schowalter, 
{\it Colloidal Dispersions} 
(Cambridge University Press, Cambridge, 1989).


\bibitem{LevinReview} 
Y. Levin, Rep. Prog. Phys. {\bf 65}, (2002) 1577.


\bibitem{Ohshima}, 
{H. Ohshima}, {\it 
Theory of Colloid and Interfacial
Electric Phenomena}, 
(Academic Press, Amsterdam, 2004).  


\bibitem{Beysens}
D. Beysens and D. Est$\grave{\rm e}$ve, 
Phys. Rev. Lett. {\bf 54}, 2123 (1985); 
 B.M. Law, J.-M. Petit, and D. Beysens, 
 Phys. Rev. E, {\bf 57}, 5782(1998); 
J.-M. Petit, B. M. Law, and D. Beysens, 
 J.Colloid Interface Sci, {\bf 202}, 441 (1998); 
D. Beysens and T. Narayanan, J. Stat. Phys. {\bf 95}, 997 (1999).

\bibitem{Maher}
P. D. Gallagher and J. V. Maher, Phys. Rev. A 46, 2012 (1992); 
P. D. Gallagher, M. L. Kurnaz, and J. V. Maher, Phys. Rev. A 46, 7750 (1992).
\bibitem{Kaler} 
Y. Jayalakshmi  and E. W. Kaler, 
Phys. Rev. Lett.  {\bf 78}, 1379 (1997).

\bibitem{Guo} 
H. Guo, T. Narayanan, M. Sztucki, 
P. Schall and G. Wegdam, Phys. Rev. Lett. {\bf 100}, 
188303 (2008)

\bibitem{Bonn}
 D. Bonn, J. Otwinowski, S. Sacanna, H. Guo, G. Wegdam and P. Schall, 
  Phys. Rev. Lett. {\bf 103}, 156101 (2009).



\bibitem{Barrat} J.L. Barrat and J.F. Joanny, 
 Adv. Chem. Phys. XCIV, I. Prigogine, S.A. Rice Eds.,
John Wiley $\&$ Sons, New York 1996.  

\bibitem{Holm} 
C. Holm,  J. F. Joanny,   K. Kremer, 
 R. R. Netz,  P. Reineker,  C. Seidel, 
T. A. Vilgis, and R. G. Winkler, 
Adv. Polym.  Sci. {\bf 166}, 67 (2004).

\bibitem{Rubinstein} A.V. Dobrynin and M. Rubinstein, 
 Prog. Polym. Sci. {\bf 30}, 1049 (2005).  


\bibitem{Evans}
P. Hopkins, A.J. Archer, and R. Evans, 
J. Chem. Phys. {\bf 131}, 124704 (2009).


\bibitem{Is} J. N. Israelachvili,  
{\it Intermolecular and Surface 
Forces} (Academic Press, London, 1991). 



\bibitem{Onuki-Kitamura} 
 A. Onuki and H. Kitamura, 
  J. Chem. Phys. {\bf 121}, 3143 (2004).
  
\bibitem{OnukiPRE} A. Onuki, Phys. Rev. E {\bf 73}, 021506 (2006); 
J. Chem. Phys.  {\bf 128}, 224704 (2008). 

\bibitem{Nara} T. Araki and A. Onuki, 
J. Phys.: Condens. Matter {\bf 21}, 424116 (2009); 
A. Onuki, T. Araki, and R. Okamoto, 
J. Phys.: Condens. Matt. {\bf 23}, 284113 (2011). 


 

\bibitem{Current} 
A. Onuki, R. Okamoto, and T. Araki, 
Bull. Chem. Soc. Jpn. {\bf 84}, 569 (2011); 
A. Onuki and  R. Okamoto, 
Current Opinion in Colloid $\&$ 
Interface Science, (Article in Press)(2011).  

\bibitem{Hung}  
Le Quoc Hung, J. Electroanal. Chem. 
{\bf 115}, 159 (1980).
\bibitem{Osakai} T. Osakai and K. Ebina, 
J. Phys. Chem. B {\bf 102}, 5691 (1998). 







\bibitem{Bu2}  I. Borukhov, D. Andelman, R. Borrega, M. Cloitre, 
L. Leibler,  and 
H. Orland, J. Phys. Chem. B  
{\bf 104}, 11027 (2000). 


\bibitem{Onuki-Okamoto} 
A. Onuki and R. Okamoto,  
J. Phys. Chem. B, {\bf 113}, 3988 (2009);  
 R. Okamoto and A. Onuki,  J. Chem. Phys. 
{\bf 131}, 094905 (2009).

\bibitem{Okamoto} 
 R. Okamoto and A. Onuki,  Phys. Rev. E {\bf 82},
 051501 (2010).  

\bibitem{Cahn} J. W. Cahn, 
J. Chem. Phys. {\bf 66} 3667 (1977).

\bibitem{Binderreview} 
K. Binder, in {\it Phase Transitions and Critical Phenomena}, 
C. Domb and J. L. Lebowitz,
eds. (Academic, London, 1983), Vol. 8, p. 1.




\bibitem{Bonnreview} D. Bonn and Ross, 
Rep.Prog.Phys.{\bf 64}, 1085 (2001).

\bibitem{h1} The adsorption becomes strong on approaching 
the criticality. At the critical composition and 
for $T>T_c$, this condition has been written as 
$|h_1| > {\rm const.}(T/T_c-1)^{\Delta_1}$ 
for neutral binary mixtures 
with the exponent $\Delta_1$ 
being estimated to be 0.5 
\cite{Fisher,Binderreview}.



\bibitem{Fisher} M.E. Fisher 
and P.G. de Gennes, 
 C. R. Acad. Sci. Paris Ser. B 287 207 (1978). 

\bibitem{Krech} M. Krech,
J. Phys.: Condens. Matt. {\bf 11}, R391 (1999). 

\bibitem{JSP} 
F. Schlesener, A. Hanke, and S. Dietrich, 
J. Stat. Phys. {\bf 110}, 981 (2003). 

\bibitem{Nature2008}
C. Hertlein, L. Helden, A. Gambassi, S. Dietrich, 
 and C. Bechinger,   Nature {\bf 451},  172 (2008). 



\bibitem{Gambassi} 
A. Gambassi, A. Macio?ek, C. Hertlein, U. Nellen, 
L. Helden, C. Bechinger, and S. Dietrich, 
Phys. Rev. E {\bf 80}, 061143 (2009). 




\bibitem{B1} 
P. G. Arscott, C. Ma, J. R. Wenner  and V. A. Bloomfield, 
Biopolymers, {\bf  36},  345 (1995).  
\bibitem{B2} 
A. Hultgren and D. C. Rau, 
Biochemistry {\bf 43}, 8272 (2004). 
\bibitem{B3} 
C.  Stanley and D. C. Rauy, 
Biophy. J.  {\bf  91}, 912 (2006). 


  
\bibitem{Sadakane} K. Sadakane, H. Seto, H. Endo, and M. Shibayama, 
J. Phys. Soc. Jpn., {\bf 76}, 113602 (2007);  
  K. Sadakane,, A.  Onuki, K.  Nishida, 
S.  Koizumi, and H.  Seto, 
Phys. Rev. Lett.  {\bf 103}, 167803 (2009);   
 K. Sadakane, N. Iguchi, 
M. Nagao, H. Endo, Y. B. Melnichenko,  and Hideki Seto, 
 Soft Matter, {\bf  7}, 1334 (2011).

\bibitem{Uchida} N. Uchida, Phys. Rev. Lett. 
{\bf 87}, 216101 (2001).  In this paper, 
the preferential adsorption is neglected, 
which is relevant for block copolymers, however. 



 \bibitem{comment-gra} 
The coefficient $C$ in  the gradient free 
energy is an arbitrary parameter in our theory. 
Using data of the surface tension, 
we could  estimate its appropriate size.   
   
\bibitem{Onukibook} A. Onuki, {\it Phase Transition Dynamics} 
(Cambridge University Press, Cambridge, 2002).

\bibitem{Tojo} K. Tojo, A. Furukawa, T. Araki, 
 and A. Onuki, 
Eur. Phys. J. E {\bf 30},  55 (2009).
The form of the electrostatic part of the free energy 
density depends on the experimental method. 


\bibitem{polar1} E.L.  Eckfeldt and 
W.W. Lucasse, 
 {J. Phys. Chem.} {\bf 47}, 164 (1943); 
B.J. Hales, G.L. Bertrand, 
 and L.G. Hepler, 
 {J. Phys. Chem.} {\bf 70}, 3970 (1966); 
V. Balevicius  and H.  Fuess, 
  Phys. Chem. Chem. Phys. {\bf{1}} ,1507 (1999).

\bibitem{largefactor} 
In the original calculation 
\cite{Okamoto}, 
the factor $\Psi (\phi_\alpha) 
= \exp[{g(\phi_\alpha-{\bar\phi})}]-1-
 g(\phi_\alpha-{\bar\phi})$ appears instead of 
 $e^{g(1-{\bar\phi})}$.  
To be precise, if  $e^{g(1-{\bar\phi})}$ 
in Eq.(4.10) is replaced by  
 $\Psi (\phi_\alpha)$, the estimated value $R_m= 35.1a_0$ 
    below Eq.(4.14) is increased to  $R_m=46.2a_0$.

 
\bibitem{touch} 
Let $\phi_\alpha$ be the 
order parameter value in the minority phase 
and $\phi_\beta$ be that of the host 
phase in the presence of a solute. 
Then the bulk precipitation curve and the 
spinodal curve touch at a point 
${\bar \phi}= \phi_{\rm pre}^{\rm cri}$ where  
$\phi_\alpha= \phi_\beta={\bar\phi}$ as in the 
right panel of Fig.4. 
Here $\phi_\alpha> \phi_\beta$ for 
${\bar \phi}< \phi_{\rm pre}^{\rm cri}$ 
and $\phi_\alpha< \phi_\beta$ for 
${\bar \phi}> \phi_{\rm pre}^{\rm cri}$. 

  
\bibitem{jump} 
  Between (C) and (D) in Fig.9, we set  
 $R_{\rm m}=46.2a_0$   in $F_{\rm wet}$ 
 in Eq.(4.15) to obtain  
$a/R_{\rm m}=0.325$ and  
$R^3-a^3=0.805R_{\rm m}^3\sim  20 a^3$ 
just after the transition, in accord with the numerical result in Fig.9. 

 

 

\bibitem{comment1} In a finite  volume $V$,  
the term proportional to $R^6/V$ generally 
appears in the droplet free energy 
$\Delta F(R)$ in  fluids.  
See Eq.(9.1.8) in Ref.\cite{Onukibook} 
for binary mixtures and 
 detailed calculations in Ref.\cite{Binder} 
  for one-component fluids. 
As a result, 
 stable  droplets 
 should have radii larger than a minimum 
 length  proportional  to  $ V^{1/(d+1)}$,  



\bibitem{Binder} 
L. G. MacDowell, P. Virnau, M.  Muller, and K. Binder, 
J. Chem. Phys. {\bf 120}, 5293 (2004). 


\bibitem{comment2} Setting $y\equiv R/R_c$, 
we rewrite Eq.(4.11) as $(R_c/R_{\rm m})^4= 
y^{-3}- y^{-4}$, so $R_c/R_{\rm m}$ decreases 
with increasing $y>4/3$. Since $y>2$ from $\Delta F>0$, 
we find  $(R_c/R_{\rm m})^4<1/16$ or  $R_c/R_{\rm m}<1/2$. 







\bibitem{Watson} G.N. Watson, 
{\it A Treatise on the Theory of Bessel Functions}, (Cambridge 
University Press, Cambridge, 1922).


\bibitem{Bessel} 
For$\ell\ge 1$ we have
$i_\ell (x) =  x^\ell ({d}/{x dx})^\ell i_0(x)$ and 
$k_\ell (x) = (- x)^\ell ({d}/{x dx})^\ell  k_0(x).$  



\end{thebibliography}
\end{document}